\RequirePackage{etoolbox}
\csdef{input@path}{{style/}{graphics/}}
\documentclass[aos,MSNbibl,dvips]{arximspdf}
\usepackage{graphicx}
\usepackage{dcolumn}

\doi{10.1214/14-AOS1276} 
\volume{43}
\issue{1}
\pubyear{2015}
\firstpage{323}
\lastpage{351}
\docsubty{FLA}

\makeatletter
\newcommand{\argmin}{\operatorname{argmin}}
\newcolumntype{d}[1]{D{(}{}{#1}}
\newtheorem{theorem}{Theorem}[section]
\newtheorem{proposition}[theorem]{Proposition}
\newcommand{\eqref}[1]{(\ref{#1})}
\makeatother

\begin{document}
\begin{frontmatter}

\title{Intermittent process analysis with scattering~moments}
\runtitle{Intermittent process analysis with scattering moments}

\begin{aug}
\author[A]{\fnms{Joan}~\snm{Bruna}\corref{}\ead[label=e1]{bruna@cims.nyu.edu}},
\author[B]{\fnms{St\'{e}phane}~\snm{Mallat}\ead[label=e2]{mallat@cmap.polytechnique.fr}\thanksref{T2}},
\author[C]{\fnms{Emmanuel}~\snm{Bacry}\ead[label=e3]{bacry@cmap.polytechnique.fr}}
\and
\author[C]{\fnms{Jean-Fran\c{c}ois}~\snm{Muzy}\ead[label=e4]{muzy@univ-corse.fr}}
\runauthor{Bruna, Mallat, Bacry and Muzy}
\affiliation{New York University,  \'Ecole Normale Sup\'{e}rieure, CNRS, Ecole Polytechnique,
and CNRS, Ecole Polytechnique and  CNRS,  Universit\'{e} de Corse}
\address[A]{J. Bruna\\
Courant Institute of Mathematical Sciences\\
Department of Computer Science \\
715 Broadway \\
New York, New York 10012\\
USA\\
\printead{e1}}
\address[B]{S. Mallat\\
Departement d'Informatique\\
\'{E}cole Normale Sup\'{e}rieure \\
45 rue d'Ulm \\
Paris 75005\\
France\\
\printead{e2}}
\address[C]{E. Bacry\\
J.-F. Muzy\\
Centre des Math\'{e}matiques Appliqu\'{e}es\\
\'{E}cole Polytechnique\\
Route de Saclay \\
Palaiseau 91128\\
France\\
\printead{e3}\\
\phantom{E-mail:\ }\printead*{e4}}
\end{aug}
\thankstext{T2}{Supported by the ANR 10-BLAN-0126 and ERC
InvariantClass 320959 grants.}

\received{\smonth{11} \syear{2013}}
\revised{\smonth{5} \syear{2014}}

\begin{abstract}
Scattering moments provide nonparametric models
of random processes with stationary increments. They are
expected values of random variables computed with a nonexpansive operator,
obtained by iteratively applying wavelet transforms and modulus nonlinearities,
which preserves the variance.
First- and second-order scattering moments are shown to characterize
intermittency and self-similarity properties of multiscale processes.
Scattering moments of Poisson processes, fractional Brownian motions,
L\'{e}vy processes and multifractal random walks are shown to
have characteristic decay.
The Generalized Method of Simulated
Moments is applied to scattering moments to estimate data generating models.
Numerical applications are shown on
financial time-series and on energy dissipation of turbulent flows.
\end{abstract}

\end{frontmatter}

\section{Introduction}

Defining nonparametric models of non-Gaussian stationary processes
remains a core issue of probability and statistics.
Computing polynomial moments is a tempting strategy
which suffers from the large variance of high order moment estimators.
Image and audio textures
are examples of complex processes with stationary increments,
which can be discriminated from a single realization by the human brain.
Yet, the amount of samples is often not sufficient
to reliably estimate polynomial moments of degree more than $2$.
These non-Gaussian processes
often have a long range dependency, and
some form of intermittency generated by randomly distributed bursts
of transient structures at multiple scales. Intermittency is an ill-defined
mathematical notion, which is used in physics to describe
those irregular bursts of large amplitude variations,
appearing, for example, in turbulent flows \cite{farge}.
Multiscale intermittency appears in other domains such
as network traffics, financial time series,
geophysical and medical data.

Intermittency is created by heavy tail processes, such as L\'{e}vy processes.
It produces large if not infinite polynomial moments of degree
larger than two, and empirical
estimations of second-order moments have a large variance.
These statistical instabilities
can be reduced by calculating expected values
of nonexpansive operators in mean-square norm, which
reduce the variance of empirical estimation.
Scattering moments are
computed with such a nonexpansive operator.
They are calculated by iteratively applying
wavelet transforms and modulus nonlinearities \cite{stephane}.
This paper shows that they characterize
self-similarity and intermittency properties of processes with
stationary increments. These properties are studied
by computing the scattering moments
of Poisson processes, fractional Brownian motions,
L\'{e}vy processes and multifractal cascades, which all have very different
behavior. Scattering moments provide nonparametric descriptors,
revealing nontrivial statistical properties of time series.
The generalized method of simulated moments
\cite{hansen,fadden} applied to scattering moments gives a parameter
estimator for data generating models, and a goodness of fit
under the appropriate statistical setting.
Besides parameter estimation,
a key challenge is to validate data generating models
from small datasets.

Section~\ref{selfsimsection} reviews the scaling properties of wavelet
polynomial moments for fractal and multifractal processes.
Scattering moments are defined and related
to multi-scale intermittency properties. Poisson processes illustrate
these first results. Section~\ref{sdfnsd8hsdf} proves that
self-similar processes with stationary increments have normalized scattering
moments which are stationary across scales.
Gaussian processes are discriminated from non-Gaussian processes using
second-order scattering moments.
Results on fractional Brownian motion and
stable L\'{e}vy processes illustrate the
multi-scale intermittency properties of these moments.
Section~\ref{multifractalscatt} extends these results to
self-similar multifractal cascades
\cite{mandelbrot69,mandelbrot1974,BKM13}.

Section~\ref{modelselection} applies scattering moments to
estimate model parameters.
It introduces a scattering moment estimator whose variance is bounded.
Parameters of data generating models
are estimated from scattering moments with
the generalized method of simulated moments \cite{hansen,fadden}.
Scattering moments of financial time-series and
turbulence energy dissipation are computed from numerical data.
Models based on
fractional Brownian, L\'{e}vy stable and multifractal cascade processes
are evaluated with a J-test.
Computations can be reproduced with
a software available at \surl{www.di.ens.fr/data/software/scatnet}.

\subsection*{Notation}
We denote $\{X(t)\}_t \stackrel{d}{=} \{Y(t)\}_t$ the equality of all
finite-\break dimensional distributions. The dyadic scaling of $X(t)$ is written
$L_j X(t) = X(2^{-j} t)$.
If $X(t)$ is stationary then $\mathbf{E}(X(t))$ does not depend on $t$ and
is written
$\mathbf{E}(X)$, and $\sigma^2(X) = \mathbf{E}(|X|^2) - \mathbf{E}(X)^2$.
We denote $B(j) \simeq F(j),  j\to\infty$ (resp., $j \to-\infty$)
if there exists $C_1,C_2>0$ and $J\in{\mathbb{Z}}$ such that
$C_1 \leq\frac{B(j)}{F(j)} \leq C_2$ for all $j>J$ (resp., for all $j
< J$).

\section{Scattering transform of intermittent processes}
\label{selfsimsection}

\subsection{Polynomial wavelet moments}
\label{wavesec}
Polynomial moments of wavelet coefficients reveal important
multiscaling properties of fractals and multi-fractals
\cite{arneoldojaffard,taqqu,multijaffard,jaffard313a,jaffard313b,abry,bacryexact91,bacry1993,WENDT2007E,WENDT2009C}.
We consider real valued
random processes $X(t)$ having stationary increments
$X(t)- X(t-\tau)$ for any $\tau\in{\mathbb{R}}$.
A wavelet $\psi(t)$ is a function of zero average
$\int\psi(t) \,dt= 0$ with $|\psi(t)| = O((1 + |t|^2)^{-1})$.
The wavelet transform of $X(t)$ at a scale $2^j$ is
defined for all $t \in{\mathbb{R}}$ by
%
\begin{equation}
\label{wavensdofis} X \star\psi_j (t)=\int X(u) \psi_j(t-u)
\,du,
\end{equation}
where $\forall j \in\mathbb{Z},\psi_j(t) = 2^{-j}\psi(2^{-j}t)$
is a
dilated version of $\psi$.
A wavelet $\psi(t)$ is said to have $q$ vanishing moments
if $\int t^k   \psi(t) = 0$ for $0 \leq k < q$.
Since $\int\psi(t)\,dt= 0$, if $X$ has stationary increments, then
one can verify that $X \star\psi_j (t)$ is a stationary process \cite{taqqu}.
The dyadic wavelet transform of $X(t)$ is
%
\begin{equation}
\label{wavetnasf}
{W}X= \{X \star\psi_j\}_{j\in\mathbb{Z}}.
\end{equation}

A wavelet $\psi$ satisfies the Littlewood--Paley condition if
its Fourier transform $\Psi$ satisfies for all $\omega\neq0$
%
\begin{equation}
\label{littlewood-paley}
\sum_{j=-\infty}^{\infty} \bigl|\Psi
\bigl(2^j \omega\bigr)\bigr|^2 + \sum
_{j=-\infty}^{\infty} \bigl|\Psi\bigl(- 2^j \omega
\bigr)\bigr|^2 = 2.
\end{equation}
If $X(t)$ is a real valued stationary process with $\mathbf{E}(|X(t)|^2) <
\infty$, then
the wavelet transform energy equals the process variance
$\sigma^2(X)$
%
\begin{equation}
\label{sdfonsdfs}
\mathbf{E}\bigl(\|{W}X\|^2\bigr) = \sum
_{j\in{\mathbb Z}} \mathbf{E}\bigl(|X \star\psi_j|^2
\bigr) = \sigma^2(X).
\end{equation}
This is proved by expressing $\mathbf{E}(|X \star\psi_j|^2)$ in terms of
the power
spectrum of $X$ and inserting (\ref{littlewood-paley}).

The decay
of monomial wavelet moments across scales can be related to the
distributions of pointwise Holder exponents
\cite{arneoldojaffard,taqqu,multijaffard,WENDT2007E,WENDT2009C}.
Moments of degree~$q$ define a scaling exponent $\zeta(q)$ such that
\[
\mathbf{E}\bigl(\bigl|X \star\psi_j (t)\bigr|^q \bigr)
\simeq2^{j \zeta(q)}.
\]
Monofractals such as fractional Brownian motions
have linear scaling exponents: $\zeta(q) = q \zeta(1)$. These Gaussian
processes have realizations which are uniformly regular.
The curvature of $\zeta(q)$ is related to
the presence of
different pointwise Holder exponents, in each realization of $X$
\cite{jaffard313a,jaffard313b,bacryexact91}. It
has been interpreted as a measurement of intermittency \cite{abry}.
Such properties cannot be obtained with Fourier moments, which
depend upon the global Holder regularity of each realization, as opposed
to pointwise Holder exponents.
However, as $q$ deviates from $1$, estimations
of moments become progressively more unstable which limits the application
of this multifractal formalism to very large data sets.
\subsection{Scattering moments}
\label{mulsdfnsc}

Scattering moments are
expected values of a nonexpansive transformation of the process.
They are computed with a
cascade of wavelet transforms and modulus nonlinearities
\cite{stephane}. We review their elementary properties.

Let $\psi$ be a $\mathbf{C}^1$, complex wavelet, whose real and
imaginary parts are orthogonal, and have the same $\mathbf{L}^2(\mathbb{R})$ norm.
In this paper, we impose that $\psi$ has a compact support
normalized to $[-1/2,1/2]$, which simplifies the proofs. However, most
results remain valid without this compact support hypothesis.
We consider wavelets~$\psi$ which are
nearly analytic, in the sense that their Fourier transform $\Psi
(\omega)$
is nearly zero for $\omega< 0$.
The compact support hypothesis prevents it from being strictly zero.
All numerical computations in the paper are performed with
the compactly supported complex wavelets of
Selesnick \cite{Selesnick}, whose real and
imaginary parts have $4$ vanishing
moments and are nearly Hilbert transform pairs.

Let $X(t)$ be a real valued process with stationary increments having
finite first-order moments: $\mathbf{E}(|X(t) - X(t-\tau)|) < \infty$
for all
$\tau\in{\mathbb R}$. The wavelet transform
$X \star\psi_{j_1}(t)$ is a complex stationary random process.
First-order scattering moments are defined by
\[
\forall j_1 \in{\mathbb{Z}}, \qquad {\overline S}X(j_1) =
\mathbf{E}\bigl(|X \star\psi_{j_1}|\bigr).
\]

First-order scattering moments do not capture the time variability
of wavelet coefficients $X \star\psi_{j_1}(t)$.
This information is partly provided by second-order
scattering moments computed from the wavelet transform of each
$|X \star\psi_{j_1}(t)|$:
\[
\forall(j_1,j_2) \in{\mathbb{Z}}^2,\qquad {
\overline S}X(j_1,j_2) = \mathbf{E}\bigl(\bigl||X \star
\psi_{j_1}| \star\psi _{j_2}\bigr|\bigr).
\]
These moments measure the average
multi-scale time variations of $|X \star\psi_{j_1}|$,
with a second family of wavelets $\psi_{j_2}$.
If $j_2 < j_1$ then
${\overline S}X(j_1,j_2)$ has a fast decay to zero as $j_1-j_2$ increases.
Its amplitudes depend on
the wavelet properties as opposed to the properties of $X$.
Indeed, if $|\psi|$ is $\mathbf{C}^p$ and has $p$ vanishing moments then
$|X \star\psi_{j_1}|$ is\vspace*{1pt} typically almost everywhere $\mathbf{C}^p$ so
${\overline S}X(j_1,j_2)=\mathbf{E}(|| X \star\psi_{j_1}| \star\psi
_{j_2}|) = O(2^{p(j_2-j_1)})$.
We thus concentrate on scattering moments for $j_2 > j_1$.

The expected value of second-order moments averages the
time variability of
$|| X \star\psi_{j_1}| \star\psi_{j_2}(t)|$. This lost information
can be recovered by calculating the wavelet transform
of $|| X \star\psi_{j_1}| \star\psi_{j_2}(t)|$ for each $(j_1,j_2)$.
Iterating this process computes
scattering moments at any order $m \geq1$:
%
\begin{equation}
\label{expansdf}
\forall(j_1,\ldots,j_m) \in{
\mathbb{Z}}^m,\qquad {\overline S}X(j_1,\ldots,j_m) =
\mathbf{E}\bigl(||X \star\psi_{j_1}| \star\cdots | \star\psi_{j_m}|\bigr).
\end{equation}
If $\mathbf{E}(|X(t) - X(t-\tau)|) < \infty$ for all $\tau\in{\mathbb
R,}$ then
$\mathbf{E}(|X \star\psi_{j_1}|) < \infty$ and
one can verify by induction on $m$ that
${\overline S}X(j_1,\ldots,j_m) < \infty$.

The vector of all scattering moments of $X$ defines a nonparametric
representation
\[
{\overline S}X = \bigl\{ {\overline S}X (j_1,\ldots,j_m) \dvtx
\forall (j_1,\ldots,j_m) \in{\mathbb Z}^m,\forall
m \in{\mathbb{N}}^* \bigr\}.
\]
Its $\ell_2$ norm is
%
\begin{equation}
\label{enfsondf}
\|{\overline S}X\|^2 = \sum
_{m=1}^{\infty} \sum_{(j_1,\ldots,j_m) \in
{\mathbb Z}^m} \bigl|{
\overline S}X (j_1,\ldots,j_m)\bigr|^2.
\end{equation}

Since the wavelet transform preserves the variance in (\ref
{sdfonsdfs}) and
the modulus operator obviously also preserves the variance,
each wavelet transform and modulus iteration preserves the variance.
If $\mathbf{E}(|X|^2) < \infty$ then by applying (\ref{sdfonsdfs}),
we verify \cite{stephane} by induction on $l$ that
scattering coefficients satisfy
\begin{eqnarray*}
&& \sum_{m=1}^{l-1} \sum
_{(j_1,\ldots,j_m) \in{\mathbb Z}^m} \bigl|{\overline S}X (j_1,\ldots,j_m)\bigr|^2
\\
&& \qquad= \sigma^2(X) - \sum_{(j_1,\ldots,j_l) \in{\mathbb Z}^l} \mathbf{E}
\bigl(||X \star\psi_{j_1}| \star\cdots | \star\psi_{j_l}|^2
\bigr),
\end{eqnarray*}
with $\sigma^2(X) = \mathbf{E}(|X|^2) - |\mathbf{E}(X)|^2$. It results that
$\|{\overline S}X\|^2 \leq\sigma^2(X)$.
Numerical experiments indicate that
for large classes of ergodic stationary processes,
$\sum_{(j_1,\ldots,j_l) \in{\mathbb Z}^l}
\mathbf{E}(||X \star\psi_{j_1}| \star\cdots | \star\psi_{j_l}|^2)$ converges
to zero as $2^l$ increases. It then implies $\|{\overline S}X\|^2 =
\sigma^2(X)$.
Similarly to the Fourier power spectrum,
the $\mathbf{l}^2 $ norm of scattering moments is then equal to the variance.
However, this remains a conjecture \cite{stephane}.

The scattering norm (\ref{enfsondf})
can be approximated with a summation restricted to moments of
order $m=1,2$, because higher order scattering moments
usually have a much smaller energy \cite{anden,pami}.
First- and second-order scattering moments applied to image and audio
textures as well as intrapartum electro-cardiograms for fetal monitoring
provide state of the art classification errors
\cite{pami,anden,sifre,CHUDACEK2013A},
but these results are strictly numerical. These algorithms are implemented
with deep convolutional neural network structures \cite{LeCun}.
In the following, we concentrate on the mathematical properties
of first- and second-order scattering moments, which characterize
self-similarity and intermittency properties.

\subsection{Normalized scattering and intermittency}
\label{sconnormals}

Scattering moments are normalized to increase their invariance properties.
Invariance to multiplicative factors is obtained with
\[
{\widetilde S}X(j_1) = \frac{{\overline S}X(j_1)} {{\overline S}X(0)} = \frac{\mathbf{E}(| X \star\psi_{j_1}|)}{
\mathbf{E}(| X \star\psi|)}.
\]
Second-order scattering moments are normalized by their first-order moment:
\[
{\widetilde S}X(j_1,j_2) = \frac{{\overline S}X(j_1,j_2)} {{\overline
S}X(j_1)} =
\frac{\mathbf{E}(|| X \star\psi_{j_1}| \star\psi_{j_2}|)}{
\mathbf{E}(| X \star\psi_{j_1}|)}.
\]
This can be rewritten
\[
{\widetilde S}X(j_1,j_2) = {\overline S}\widetilde X_{j_1} (j_2) = \mathbf{E}\bigl(| \widetilde X_{j_1}
\star\psi_{j_2}|\bigr) \qquad\mbox{with }\widetilde X_{j_1} =
\frac{|X \star\psi_{j_1}|}{
\mathbf{E}(| X \star\psi_{j_1}|)}.
\]
If $X$ has stationary increments then $\widetilde X_{j_1}$ is a normalized
stationary
process providing the occurrence of ``burst'' of activity at the scale
$2^{j_1}$. Normalized second-order moments ${\widetilde S}X(j_1,j_2)$
thus measure
the time variability
of these burst of activity over time scales $2^{j_2} \geq2^{j_1}$,
which gives multi-scale measurements of intermittency.

Intermittency aims at capturing the geometric distribution
of burst of high variability in each realization of $X$.
It is not modified by the action
of derivative operators, which are translation invariant.
We verify that this invariance property holds
for normalized second-order moments. Let
$d^\alpha$ be a fractional derivative
defined by the multiplication by $(i \omega)^\alpha$ in the
Fourier domain. Since
\[
d^\alpha X \star\psi_{j_1}(t) = 2^{-\alpha j_1} X \star\psi
^\alpha_{j_1} (t),
\]
where $\psi^\alpha= d^\alpha\psi$ and $\psi^\alpha_{j_1} (t) =
2^{-j_1} \psi^\alpha(2^{-j_1} t)$, it results that
%
\begin{equation}
\label{ndf08sd}
{\overline S}d^\alpha X (j_1) =
2^{-\alpha j_1} \mathbf{E}\bigl(\bigl|X \star\psi ^\alpha_{j_1}\bigr|
\bigr)
\end{equation}
and
%
\begin{equation}
\label{ndf08sdd} {\widetilde S}d^\alpha X(j_1,j_2)
= \frac{\mathbf{E}(|| X \star\psi^\alpha_{j_1}| \star\psi_{j_2}|)}{
\mathbf{E}(| X \star\psi^\alpha_{j_1}|)}.
\end{equation}
If $X(t)$ has no oscillating singularity \cite{jaffardMeyer},
then its wavelet coefficients
calculated with $\psi$ and $\psi^\alpha$ have the same asymptotic
decay, so
%
\begin{equation}
\label{ndf08sdd3}
{\overline S}d^\alpha X (j_1)
\simeq2^{-\alpha j_1} {\overline S}X(j_1)\quad\mbox{and}\quad {\widetilde S}d^\alpha X(j_1,j_2) \simeq{\widetilde S}X
(j_1,j_2).
\end{equation}
Modifications of regularity produced by derivative operators
affect the decay of
first-order scattering moments but not the
decay of normalized second-order moments.
Fractional Brownian motions illustrate these properties
in Section~\ref{gaussianwhitesection}.

Global intermittency parameters computed with wavelet moments
can be related to normalized second-order scattering moments.
If $\mathbf{E}(|X \star\psi_{j}|^q) \simeq2^{j \zeta(q)}$ then
intermittency is measured by the curvature of $\zeta(q)$.
If one quantifies it by $\zeta(2) - 2 \zeta(1)$, we verify
from (\ref{sdfonsdfs}) that
\[
2^{j(\zeta(2) - 2 \zeta(1))} \simeq \frac{\mathbf{E}(|X \star\psi_{j}|^2)}{
\mathbf{E}(|X \star\psi_{j}|)^2} \geq1 + \sum
_{j_2 = -\infty
}^{+\infty} \bigl|{\widetilde S}X(j,j_2)\bigr|^2.
\]

%
%
%

It results
that if $\sum_{j_2 = -\infty}^{+\infty} {\widetilde S}X(j,j_2)^2
\simeq
2^{j \beta}$ as $j \to-\infty$ with $\beta<0$, then $\zeta(2) - 2
\zeta(1) < 0$.
However, these moments eliminate the dependency
on the scale parameter $2^{j_2}$, which
provides a finer multi-scale characterization of the intermittency regularity.
This dependency upon $2^{j_2}$ is studied in the next sections and is used
for model selection in Section~\ref{modelselection}.

\begin{figure}[b]

\includegraphics{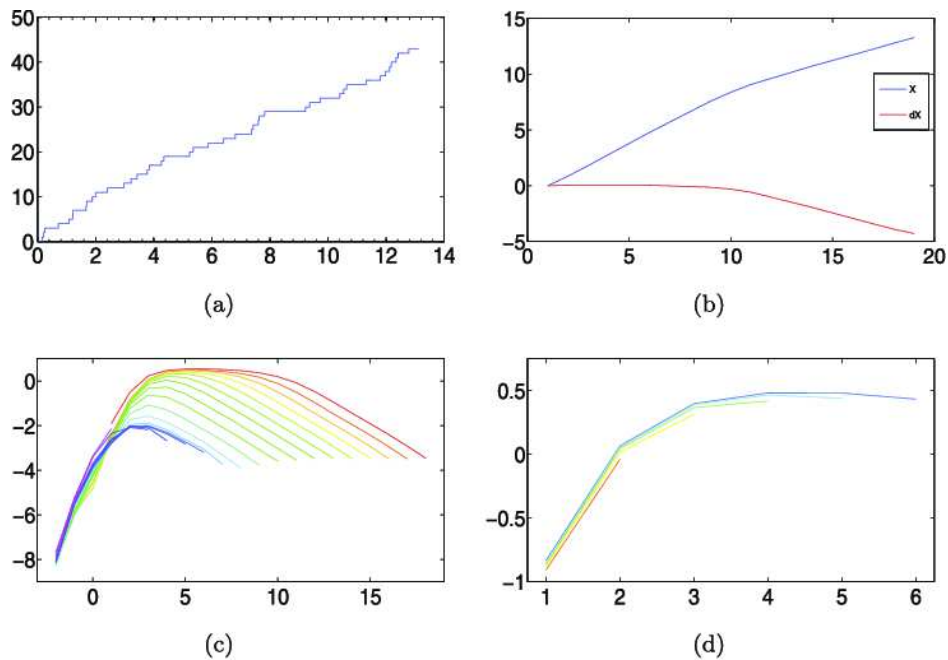}
\vspace*{-3pt}
\caption{\textup{(a)} Realization of a Poisson process $X(t)$ of intensity $\lambda= 10^{-4}$.
\textup{(b)} $\log_2{\widetilde S}X(j)$ and $\log_2{\widetilde S}\,dX(j)$ as a
function of $j$.
\textup{(c)} $\log_2{\widetilde S}X(j_1,j_2)$ as a function of $j_2 - j_1$ for
several values of $j_1$.
\textup{(d)} The same curves as in \textup{(c)}, but restricted to $j_2 < -\log
_2(\lambda)-1$.}
\label{poisson-figure}
\vspace*{-6pt}
\end{figure}

\subsection{Scattering Poisson processes}

The properties of scattering moments are illustrated over a Poisson
process, which is a simple L\'evy process with stationary increments.
A homogeneous Poisson process $\{X(t), t\geq0\}$ has increments
$X(t+\Delta) - X(t)$ which count the
number of occurrence of events in $]t,t+\Delta]$,
and have a Poisson distribution of intensity $\lambda$.
Figure~\ref{poisson-figure}(a) shows an example.
The following proposition gives the decay
of first- and second-order scattering moments of Poisson\vspace*{-3pt} processes.

\begin{theorem}
\label{poissonprop}
If $X$ is a Poisson process of intensity $\lambda$ and
$\bar\psi(t) = \int_0^t \psi(u)\,du$ then for all $j_1 \leq j_2$
%
\begin{eqnarray}
\label{poissonfirst}
{\overline S}X (j_1) &=&  2^{j_1} \lambda\|
\bar\psi\|_1 \bigl(1 + O\bigl(2^{j_1} \lambda\bigr) \bigr),
\\
\label{poissonfirstbis}
\lim_{j_1 \to\infty} 2^{-j_1/2} {\overline
S}X(j_1)& =&  C \lambda ^{1/2} > 0,
\end{eqnarray}
where $C$ depends only upon the wavelet $\psi$, and
%
\begin{eqnarray}
\label{poissonsec}
\widetilde{S} X (j_1, j_2) &=&
\frac{\| |\bar\psi| \star\psi_{j_2-j_1} \|_1} {\|\bar\psi\|_1} \bigl(1 + O\bigl(\lambda2^{j_1}\bigr) + O\bigl(
\lambda2^{j_2} \bigr) \bigr),
\\
\label{poissonsecbis}
\lim_{j_2 \rightarrow\infty} 2^{j_2/2} \widetilde{S} X
(j_1, j_2) &=& C' > 0.
\end{eqnarray}
\end{theorem}

The proof is in Appendix~A in \cite{suppA}.
At scales $2^{j_1} \leq2^{j_2} \ll\lambda^{-1}$,
the Poisson process typically has 1 jump over the support of
each wavelet, which implies
(\ref{poissonfirst}).
When $2^{j_1} \gg\lambda^{-1}$,
Appendix~A proves that $X \star\psi_{j_1} (t)$
converges\vspace*{1pt} to the wavelet transform of a Gaussian white noise of
variance $\lambda 2^{j_1}$, which implies~(\ref{poissonfirstbis}).

When $2^{j_2} \ll\lambda^{-1}$,
(\ref{poissonsec}) implies that
\[
\lim_{j_1 \to-\infty} \widetilde S X(j_1,j_2) =
\|\psi\|_1 \bigl(1 + O\bigl(2^{j_2} \lambda\bigr) \bigr).
\]
This convergence to a constant indicates a high degree of intermittency,
because fine scale wavelets see individual Diracs occurring randomly.
This property is observed in
Figure~\ref{poisson-figure}(d), which gives
$\log_2 \widetilde S X(j_2-j_1,j_2)$ as a function of $j_2-j_1$.
These curves overlap
for different $j_1$, and converge to $\|\psi\|_1$.

If $2^ {j_2} \gg\lambda^{-1}$
then $\widetilde S X(j_1,j_2) \simeq 2^{-j_2/2}$.
This decay is characteristic of Gaussian stationary processes,
which are uniformly regular,
and thus have no intermittency. This is further studied
in Section~\ref{gaussianwhitesection} for fractional Brownian motions.
Figure~\ref{poisson-figure}(c) verifies that
$\widetilde S X(j_2-j_1,j_2)$ decays with a slope of $-1/2$ as a function
of $j_2-j_1$.

When going from $X$ to $dX$ then the sum of jumps is replaced by a measure
which is a sum of Diracs. We
verify from Appendix~A that
${\overline S}\,dX(j_1) \simeq2^{-j_1} {\overline S}X(j_1)$. This\vspace*{1pt}
reflects the change\vspace*{1pt} of regularity.
Figure~\ref{poisson-figure}(b) shows that the
difference between the slopes of $\log_2\widetilde S X(j_1)$ and\vspace*{1pt}
$\log_2\widetilde S \,dX(j_1)$ is indeed equal to $1$.
For normalized second-order moments,
$\widetilde S \,dX(j_1,j_2)$ is nearly equal to $\widetilde S X(j_1,j_2)$,
as expected from (\ref{ndf08sdd}).

\section{Self-similar processes}
\label{sdfnsd8hsdf}
Second-order scattering moments of self-similar processes are proved to
be stationary across scales.
Fractional Brownian
motions and L\'evy stable processes are studied in Sections~\ref{gaussianwhitesection} and \ref{Levy}.

\subsection{Scattering self-similarity}
Self-similar processes of Hurst exponent $H$
are stochastic processes $X(t)$ which are invariant in
distribution under a scaling of space or time:
%
\begin{equation}
\label{selfsim}
\forall s >0,\qquad\bigl\{ X(st) \bigr\}_t \stackrel{d} {=}
\bigl\{ s^H X(t) \bigr\}_t.
\end{equation}
We consider self-similar processes having stationary increments.
Fractional
Brownian motions and $\alpha$-stable L\'{e}vy processes are examples
of Gaussian and non-Gaussian self-similar processes with stationary increments.

If $X$ is self-similar, then applying (\ref{selfsim}) with
a change of variable $u' = 2^{-j} u$ in~(\ref{wavensdofis}) proves that
\[
\forall j \in\mathbb{Z},\qquad\bigl\{X \star\psi_j(t)\bigr\}_t
\stackrel{d} {=} 2^{jH } \bigl\{ X \star\psi\bigl(2^{-j} t
\bigr)\bigr\}_t.
\]
The following proposition proves that normalized second-order
scattering moments can be written
as a univariate function.

\begin{proposition}
\label{self-sim-prop}
If $X$ is a self-similar process with stationary increments then
for all $j_1 \in{\mathbb{Z}}$
%
\begin{equation}
\label{expansdf3}
{\widetilde S}X(j_1) = 2^{{j_1}H},
\end{equation}
and for all $(j_1,j_2) \in{\mathbb{Z}}^2$
%
\begin{equation}
\label{pasfns}
{\widetilde S}X (j_1,j_2) = {\overline S}
{\widetilde X}(j_2-j_1)\qquad\mbox{with } {\widetilde X}(t) =
\frac{|X \star\psi(t)|} {\mathbf{E}(|X
\star\psi|)}.
\end{equation}
\end{proposition}

\begin{pf}
We write $L_j x(t) = x(2^{-j} t)$.
Since $\psi_{j_1} = 2^{-{j_1}} L_{j_1} \psi$,
a change of variables yields
$L_{j_1} |X \star\psi| = | L_{{j_1}} X \star\psi_{j_1} |$,
and hence
%
\begin{equation}
\label{expansdf0}
|X \star\psi_{j_1} | = L_{j_1} |
L_{-{j_1}} X \star\psi| \stackrel{d} {=} 2^{{j_1}H}
L_{j_1}|X \star\psi|.
\end{equation}
If $Y(t)$ is stationary, then $\mathbf{E}(L_j Y(t))=\mathbf{E}(Y(t))$, which proves
(\ref{expansdf3}).

By cascading (\ref{expansdf0}), we get
%
\begin{equation}
\label{sdf0}
\forall (j_1,j_2),\qquad \bigl||X \star
\psi_{j_1}|\star\psi_{j_2}\bigr| \stackrel{d} {=} 2^{j_1 H}
L_{j_1}\bigl|| X \star\psi| \star\psi_{j_2 - j_1}\bigr|,
\end{equation}
so ${\overline S}X(j_1,j_2) = 2^{j_1 H} \mathbf{E}(||X \star\psi| \star
\psi_{j_2-j_1}|)$.
Together with (\ref{expansdf3}), it proves (\ref{pasfns}).
\end{pf}

\begin{figure}

\includegraphics{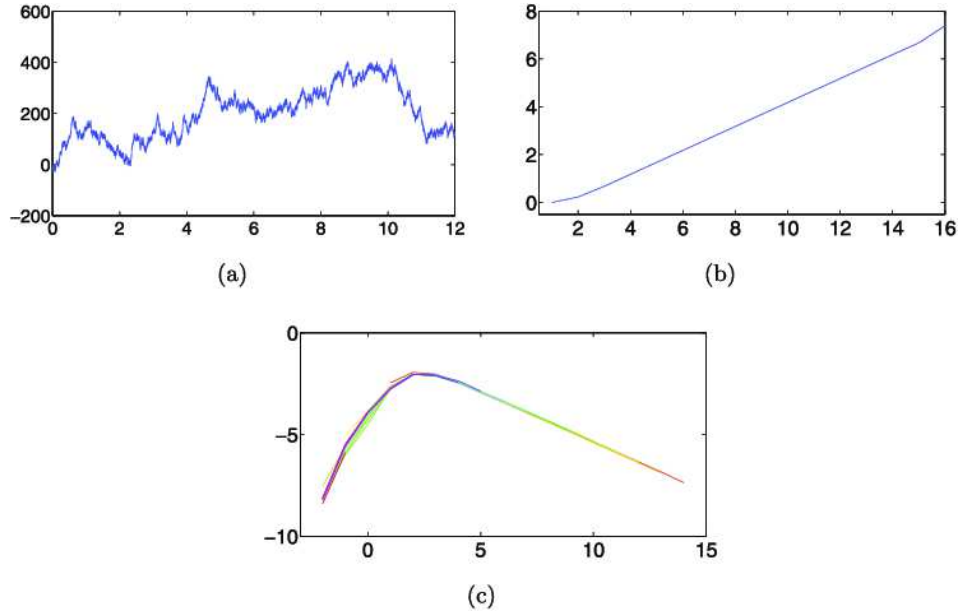}

\caption{\textup{(a)} Realization of a Brownian motion $X(t)$.
\textup{(b)} $\log_2 {\widetilde S}X(j_1)$ as a function of $j_1$.
\textup{(c)} The curves $\log_2 {\widetilde S}X(j_1,j_1+l)$ as a function of $l$
are identical for different $j_1$.}
\label{brownian-figure}
\end{figure}

Property (\ref{pasfns}) proves that if $X$ is self-similar then
${\widetilde S}X(j_1,j_1+l)$
is a function of~$l$, which can be interpreted as a stationary
property across scales. This function of $l$ is a scattering
intermittency measure of the random process.
A Brownian motion
is a Gaussian self-similar process with a Hurst exponent $H = 1/2$.
It results from (\ref{expansdf3})
that $\log_2 {\widetilde S}(j_1) = j_1/2$,
which is illustrated by Figure~\ref{brownian-figure}(b).
Figure~\ref{brownian-figure}(c) displays
${\widetilde S}X(j_1,j_2)$ expressed as a function of
$j_2-j_1$, for different $j_1$.\vspace*{1pt} The curves for different $j_1$ are
equal, as proved by (\ref{expansdf3}).
When $j_2 - j_1 < 0$, ${\widetilde S}X(j_1,j_2)$
increases with a slope which
does not depend on $X$ but on the number of vanishing moments and on
the regularity of the wavelet $\psi$.
For $j_2 - j_1 \geq0$,\vspace*{1pt} the decay depends upon the property of $X$ and
satisfies
${\widetilde S}X(j_1,j_2) \simeq2^{-(j_2-j_1)/2}$. The next section
proves this result
in the more general context of fractional Brownian motions, and shows that
it reflects the fact that a Brownian motion is a Gaussian process.

\subsection{Fractional Brownian motions}
\label{gaussianwhitesection}

We compute the normalized scattering
representation of
fractional Brownian motions, which are
the only self-similar Gaussian processes with stationary increments.
A fractional Brownian motion of Hurst exponent $0<H<1$ is defined
as a zero mean Gaussian process $\{X(t)\}$, satisfying
\[
\forall t,s>0,\qquad \mathbf{E}\bigl(X(t) X(s)\bigr) = \tfrac{1} 2
\bigl(t^{2H} + s^{2H} - |t-s|^{2H} \bigr) \mathbf{E}
\bigl(X(1)^2\bigr).
\]
It is self-similar and satisfies
\[
\forall s>0,\qquad \bigl\{X(st)\bigr\}_t \stackrel{d} {=} s^H
\bigl\{X(t)\bigr\}_t.
\]

\begin{figure}

\includegraphics{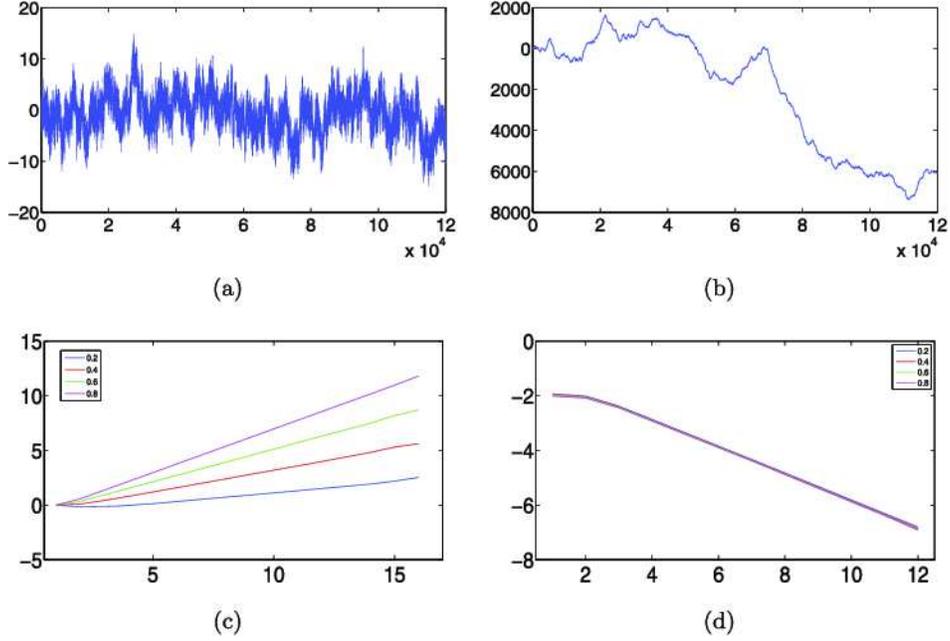}

\caption{\textup{(a)}, \textup{(b)} Realizations of fractional Brownian motions $X(t)$
with $H=0.2$ in \textup{(a)} and $H = 0.8$ in \textup{(b)}.
\textup{(c)} $\log_2 {\widetilde S}X (j_1)$ as a function of $j_1$, for
$H = 0.2,  0.4, 0.6,   0.8$. Slopes are equal to $H$.
\textup{(d)}~$\log_2 \widetilde S X (j_1,j_1+l)$
as a function of $l$ do not depend on $j_1$ for all $H$.}
\label{a-fbm}
\end{figure}

Proposition~\ref{self-sim-prop} proves in
(\ref{expansdf3}) that
${\widetilde S}X(j_1) = 2^{H j_1}$.
This is verified by Figure~\ref{a-fbm}(a) which shows that
$\log_2 {\widetilde S}X(j_1) = H j_1$ for several
fractional Brownian motions with $H=0.2,0.4,0.6,0.8$.

Figure~\ref{a-fbm}(c)
displays
$\log_2 {\widetilde S}(j_1,j_2)$, which is a function of $j_2 - j_1$,
as proved
by~(\ref{pasfns}).
Modulo a proper initialization at $t = 0$, if $X$ is
a fractional Brownian motion of exponent $H$ then $d^\alpha X$ is a fractional
Brownian motion of exponent $H-\alpha$.
We thus expect from (\ref{ndf08sdd3})
that $\log_2 {\widetilde S}X(j_2-j_1)$ nearly
does not depend upon~$H$. This is shown by
Figure~\ref{a-fbm}(c) where all curve superimpose for $j_2 - j_1 > 0$,
with a slope of $-1/2$. This result is proved by the following theorem.

\begin{theorem}
\label{fbmpropo}
Let $X(t)$ be a fractional Brownian motion with
Hurst exponent $0 < H < 1$. There exists a constant $C>0$ such that
for all $j_1 \in{\mathbb{Z}}$
%
\begin{equation}
\label{brownianresult}
\lim_{l \to\infty} 2^{l/2} {\widetilde S}X
(j_1,j_1+l) = C. 
\end{equation}
\end{theorem}

For a fractional Brownian motion,
$\log_2 \widetilde S X(j_1,j_1+l)$ does not depend
on $j_1$ or $H$, and its slope is thus equal to $-1/2$
when $l$ increases. This value is characteristic of
wide-band Gaussian stationary processes. It indicates that there is no
intermittency phenomenon at any\ scale.

\subsection{\texorpdfstring{$\alpha$}{alpha}-stable L\'{e}vy processes}
\label{Levy}

In this section, we compute the scattering moments
of $\alpha$-stable L\'{e}vy processes and analyze their intermittency behavior
for $1 < \alpha\leq2$.
These processes have finite polynomial moments only for
degree strictly smaller than $\alpha\leq2$ \cite{levytutorial}.
Indeed, for $\alpha> 1$, $\alpha$-stable L\'evy process
$X(t)$ have stationary increments and
$\mathbf{E}(|X(t)-X(t-\tau)|) < \infty$ for any $\tau\in{\mathbb R}$.
Its scattering moments are thus well defined at all orders.
Self-similar L\'{e}vy processes have stationary increments with
heavy tailed
distributions. Their realizations contain rare, large
jumps, which are responsible
for the blow up of moments larger than $\alpha$.
They induce a strongly intermittency behavior.

\begin{figure}

\includegraphics{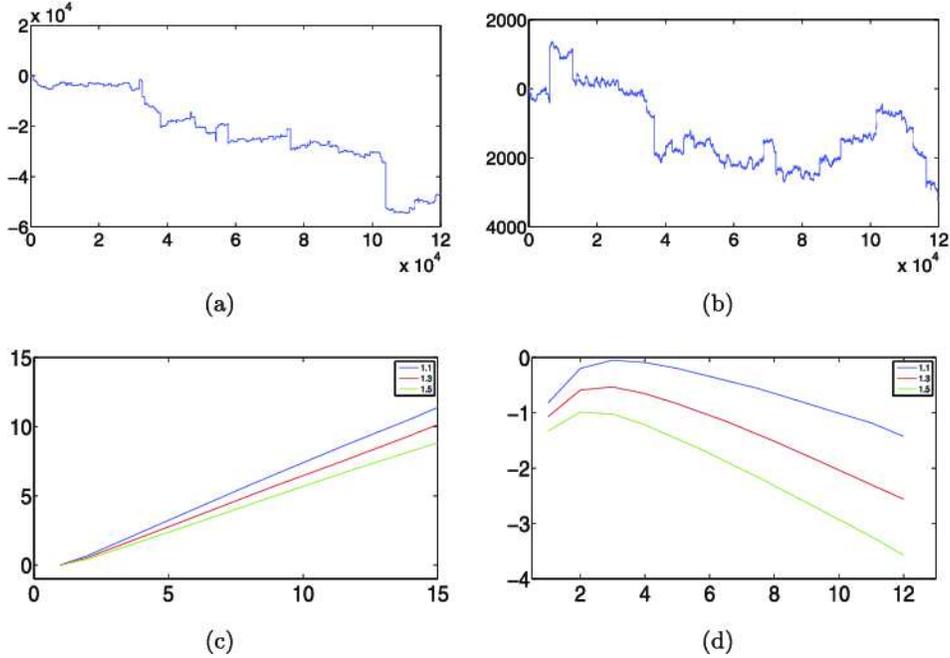}

\caption{\textup{(a)}, \textup{(b)} Realizations of $\alpha$-stable L\'{e}vy processes $X(t)$
with $\alpha=1.1$
and $\alpha=1.5$.
\textup{(c)} $\log_2 {\widetilde S}X_\alpha(j_1)$ as a function of $j_1$
with $\alpha= 1.1,   1.2,   1.3$. Slopes are equal to $\alpha^{-1}$.
\textup{(d)}~$\log_2 {\overline S}X_\alpha(j_1,j_1 +l)$ as a function of $l$ do
not depend on $j_1$. Slopes tend to $\alpha^{-1}-1$ when $l$ increases.}
\label{levyscatteringfigure}
\end{figure}

$X(t)$ satisfies the self-similarity relation
%
\begin{equation}
\label{expansdf3levysdf}
\bigl\{X (st)\bigr\}_t \stackrel{d} {=}
s^{\alpha^{-1}} \bigl\{X(t)\bigr\}_t,
\end{equation}
so Proposition~\ref{self-sim-prop} proves that
%
\begin{equation}
\label{expansdf3levy}
{\widetilde S}X(j_1) = 2^{{j_1}\alpha^{-1}}.
\end{equation}
This is verified in Figure~\ref{levyscatteringfigure} which shows that
$\log_2 {\widetilde S}X(j_1) = \alpha^{-1} j_1$.
First-order moments
thus do not differentiate a L\'{e}vy stable processes from
fractional Brownian motions of Hurst exponent $H = \alpha^{-1}$.

The self-similarity implies that
${\widetilde S}X (j_1,j_1+l)$ does not depend on $j_1$. However, they
have a very
different behavior than second-order scattering moments of fractional
Brownian motion. Figure~\ref{levyscatteringfigure} shows
that $\log_2 {\widetilde S}X(j)$ has a slope which tends to $\alpha
^{-1}-1$ and hence that
when $l$ increases
%
\begin{equation}
\label{ldevnoudsfY}
{\widetilde S}X (j_1,j_1+l)
\simeq2^{l(\alpha^{-1}-1)}.
\end{equation}
For $\alpha< 2$, then $\alpha^{-1}-1 > -1/2$, so ${\widetilde
S}X(j_1,j_1+l)$ has
a slower decay for $\alpha$-stable L\'evy processes than for fractional
Brownian motion, which corresponds to the fact that these processes
are highly intermittent and the intermittency increases when $\alpha$
decreases.
For $\alpha= 2$, the L\'{e}vy process $X$ is a Brownian motion and we
recover that ${\widetilde S}X(j_1,j_1+l) \simeq2^{-l/2}$ as proved in
Theorem~\ref{fbmpropo}.

The scaling property (\ref{ldevnoudsfY}) is
explained qualitatively, without formal proof.
We proved in (\ref{pasfns}) that
%
\begin{equation}
\label{ldevnoudsfYsdf}
{\widetilde S}X (j_1,j_2) =
\frac{\mathbf{E}(||X \star\psi(t)| \star
\psi_{l}|)} {\mathbf{E}(|X \star\psi|)}\qquad\mbox{for }l = j_2-j_1.
\end{equation}
The stationary process $|X \star\psi(t)|$ measures the amplitude of
local variations of the process $X$. It is dominated by
a sparse sum of large amplitude
bumps of the form $a |\psi(t-u)|$,
where $a$ and $u$ are the random amplitudes and positions of
rare large amplitude jumps in $X(t)$, distributed according to the L\'
{e}vy measure. It results that
%
\begin{equation}
\label{ldevnoudsfYsdf0}
\mathbf{E}\bigl(\bigl||X \star\psi| \star\psi_l\bigr|\bigr) \simeq
\mathbf{E}\bigl(\bigl|dX \star|\bar\psi| \star\psi_l\bigr|\bigr)\qquad\mbox{with }\overline
\psi(t) = \int_0^t \psi(u)\,du.
\end{equation}
If $2^l \gg1$, then $|\bar\psi| \star\psi_l \approx\|\bar\psi\|
_1  \psi_l$,
and $\mathbf{E}(|dX \star\psi_l|) \simeq2^{l(\alpha^{-1}-1)}$
because the L\'evy measure $dX(t)$ satisfies the self-similarity property
\[
\bigl\{dX (st)\bigr\}_t \stackrel{d} {=} s^{\alpha^{-1}-1} \bigl
\{dX(t)\bigr\}_t.
\]
Inserting (\ref{ldevnoudsfYsdf0}) in
(\ref{ldevnoudsfYsdf}) gives the scaling property (\ref{ldevnoudsfY}).

\section{Scattering moments of multiplicative cascades}
\label{multifractalscatt}

We study the scattering representation of multifractal processes
which satisfy a stochastic scale invariance property.
Section~\ref{scattMRMsection}
studies the particularly important case of
log-infinitely divisible multiplicative processes.

\subsection{Stochastic self-similar processes}
\label{lognsdf8hsdf}

We consider processes with stationary increments which satisfy
the following stochastic self-similarity:
%
\begin{equation}
\label{selfsim2}
\forall 1 \geq s >0,\qquad \bigl\{X(st)\bigr\}_{t \leq2^L}
\stackrel{d} {=} A_s \cdot\bigl\{X(t)\bigr\}_{t\leq2^L},
\end{equation}
where $A_s$ is a log-infinitely\vspace*{1pt} divisible random variable independent
of $X(t)$ and the so-called \textit{integral scale}
$2^L$ is chosen (for simplicity) as a power of $2$.
The Multifractal Random Measures (MRM) introduced by \cite{MB02,BM03}
are important examples of such processes. Let us point out that MRMs are
stationary increments versions of grid bound
multiplicative cascades initially introduced by Yaglom \cite{yaglom66}
and Mandelbrot \cite{mandelbrot69,mandelbrot1974},
and further studied by Kahane and Peyriere \cite{kahane}. In that respect,
all the results that we obtain on MRMs can be easily generalized to
discrete multiplicative cascades.
For the sake of conciseness, we did not include them here.

Since $X$ has stationary
increments and satisfies (\ref{selfsim2}),
with a change of variables, we verify that $\forall j \leq L$, $\{ X
\star\psi_j(t)\}_t
\stackrel{d}{=} A_{2^j}  \{ X \star\psi(2^{-j} t)\}_t$,
and hence, for all $q \in{\mathbb{Z}}$ and $j \leq L$
%
\begin{equation}
\label{scalingexpowave}
\mathbf{E}\bigl(| X \star\psi_j|^q\bigr)
= \mathbf{E}\bigl(|A_{2^j}|^q\bigr) \mathbf{E}\bigl\{|X
\star\psi|^q\bigr\}\simeq C_q 2^{j \zeta(q)},
\end{equation}
where $\zeta(q)$ is a priori a nonlinear concave function of $q$
\cite{jaffard313a,jaffard313b}.
Similarly to Proposition~\ref{self-sim-prop},
the following proposition shows that
normalized scattering moments capture stochastic
self-similarity with a univariate function.

\begin{proposition}
\label{self-sim-prop2}
If $X$ is randomly self-similar in the sense of (\ref{selfsim2})
with stationary increments, then
for all $j_1 \leq L$
%
\begin{equation}
\label{expansdf30} {\widetilde S}X (j_1) = \mathbf{E}\bigl(|A_{2^{j_1}}|\bigr)
.
\end{equation}
Moreover, if $2^{j_1}+2^{j_2} \leq L$ then
%
\begin{equation}
\label{pasfns0}
{\widetilde S}X (j_1,j_2) = {\overline S}
{\widetilde X}({j_2-j_1})\qquad\mbox{with } {\widetilde X}(t) =
\frac{|X \star\psi
(t)|} {\mathbf{E}(|X \star\psi|)}.
\end{equation}
\end{proposition}

\begin{pf}
Property (\ref{expansdf3}) is a particular case of (\ref{scalingexpowave})
for $q = 1$.
If $j_1 + j_2 \leq L$, with the same derivations as for
(\ref{sdf0}), we derive from (\ref{selfsim2}) that
%
\begin{equation}
\bigl||X \star\psi_{j_1}|\star\psi_{j_2}\bigr| \stackrel{d} {=}
A_{2^{j_1}} L_{j_1}\bigl|| X \star\psi| \star\psi_{j_2
- j_1}\bigr|,
\end{equation}
so ${\overline S}X(j_1,j_2) = \mathbf{E}(A_{2^{j_1}}) \mathbf{E}(||X \star
\psi| \star\psi_{j_2-j_1}|)$.
Together with (\ref{expansdf30}), it proves (\ref{pasfns0}).
\end{pf}

\begin{figure}

\includegraphics{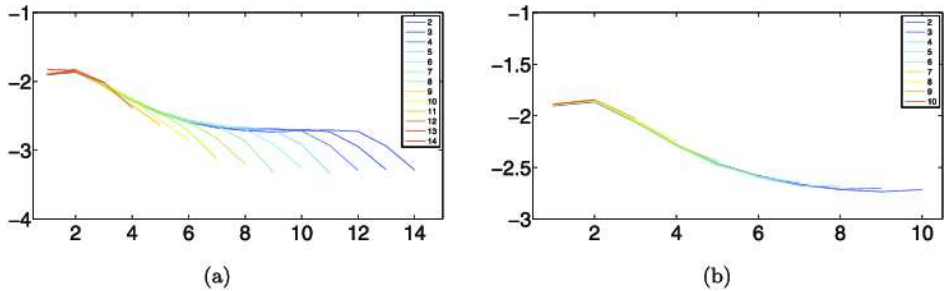}

\caption{\textup{(a)} $\log_2 {\overline S}X_\alpha(j_1,j_1 +l)$ as a
function of $l$
for a Multifractal Random Measure (MRM) with $\lambda^2=0.04$ and
an integral scale $2^L = 2^{13}$.
Different colors stand for different values of $j_1$.
\textup{(b)} Same curves restricted to $j_2 = j_1 + l < L-1$.}
\label{MRMfulltransferfig}
\end{figure}

Figure~\ref{MRMfulltransferfig}\vspace*{1pt} shows the
normalized scattering of a multiplicative cascade process
described in Section~\ref{scattMRMsection},
with an integral scale $2^L = 2^{17}$.
When $2^{j_2} \geq2^L$ is beyond the integral scale,
as for a Poisson process, wavelet\vspace*{1pt} coefficients converge
to Gaussian processes. It results that $\log_2 {\widetilde
S}(j_1,j_2)$ decays
with a slope $-1/2$ as a function of $j_2-j_1$ for $j_2 > L$, as
shown in Figure~\ref{MRMfulltransferfig}(a).
If $j_1 < j_2 < L$ then
(\ref{pasfns0}) proves that ${\widetilde S}X (j_1,j_2)$ only depends
on $j_2-j_1$,
and all curves in Figure~\ref{MRMfulltransferfig}(b) superimpose in
this range.

Propositions \ref{self-sim-prop} and \ref{self-sim-prop2} show
that the stationary property
${\widetilde S}X(j_1,j_2)={\widetilde S}X(j_2-j_1)$
can be used to
detect the presence of self-similarity, both deterministic
and stochastic. This necessary
condition is an alternative to the
scaling of the $q$-order moments,
$\mathbf{E}(|X \star\psi_j|^q) \simeq C_q 2^{j\zeta(q)}$,
which is difficult to verify empirically for $q\geq2$
or $q<0$.


\subsection{Log-infinitely divisible Multifractal Random Processes}
\label{scattMRMsection}
%

Multiplicative cascades as introduced by Mandelbrot in \cite
{mandelbrot69,mandelbrot1974} are built as an iterative process
starting at scale $2^L$. They are obtained as the (weak) limit of the
measure $dM_n$ whose restriction
over a dyadic interval of the form $[k2^{L-n},(k+1)2^{L-n}]$ is uniform
and equal to $\prod_{i=1}^{n} W_{i}^{(k)} \,dt$,
where the $W_{i}^{(k)}$'s are i.i.d. log-infinitely divisible random variables.
Multifractal Random Measures (MRM), introduced in \cite{MB02,BM03},
are the stationary increments versions of these multiplicative cascades,
and are an important class of processes reproducing multi-fractal
behavior while having stationary increments.
They are built using an infinitely divisible random noise $dP$
distributed in the half-plane $(t,s)$ ($s>0$). Using the previous
notations, the noise around $(t,s)$ can be seen as the equivalent of
the infinitely divisible variable
$\log_2 W_{\log s}^{(t/s)}$.
More precisely, if $\omega_l^{2^L}(t) = \int_{\mathcal{A}_l^{2^L}(t)}
\,dP$ where $\mathcal{A}_l^{2^L}(t)$ is the cone in the $(t,s)$ half-plane
pointing to point $(t,0)$ and truncated for $s<l$,
the MRM is defined as the weak limit: $dM(t) = \lim_{l \rightarrow0}
e^{\omega_l^{2^L}(t)} \,dt$.
For a rigorous definition of $\omega_l^{2^L}$ and of a Multifractal
Random Measure, we refer the reader to \cite{BM03}.

One can prove that $dM $ satisfies \eqref{selfsim2}, and hence that is
multifractal, in the sense that
\[
\mathbf{E}\bigl(| X \star\psi_j|^q\bigr) = \mathbf{E}
\bigl(|A_{2^j}|^q\bigr) \mathbf{E}\bigl\{|X \star
\psi|^q\bigr\}\simeq C_q 2^{j \zeta(q)},
\]
where $\zeta(q)$ is a nonlinear function
which is uniquely defined by the infinitely divisible law chosen for $dP$.
If $dP$ is Gaussian, $dM$ is generally referred to as a ``log-Normal"
MRM, and in this case \cite{BM03}:
%
\begin{equation}
\label{valuezeta} \zeta(q) = \biggl(1 + \frac{\lambda^2} 2\biggr) q -
\frac{\lambda^2} 2 q^2.
\end{equation}
The curvature of the concave function
$\zeta(q)$ at $q=0$ ($\lambda^2$ in the latter case) plays the role
of the so-called ``intermittency factor'' in
the multifractal formalism \cite{jaffard313a,jaffard313b}. The larger $\lambda^2$,
the more intermittency.

The self-similarity properties of $dM$ are mainly direct consequences
of the ``global'' self-similarity properties of $\omega^{2^L}_l$:
%
\begin{equation}
\label{elaw} \bigl\{ \omega_{sl}^{s2^L}(st)\bigr
\}_t  \mathop{=}^{\mathrm{law}}\bigl\{\omega _l^{2^L}(t)
\bigr\}_t, \qquad\forall L, \forall s>0,
\end{equation}
and of the stochastic self-similarity property:
%
\begin{equation}
\label{elaw2} \bigl\{ \omega_{sl}^{2^L}(su)\bigr
\}_{u<T} \mathop{=}^{\mathrm{law}}  \bigl\{ \Omega_s +
\omega_l^{2^L}(u)\bigr\}_{u<T},\qquad\forall L,\forall s
<1,
\end{equation}
where $\Omega_s$ is an infinitely divisible random variable
independent of $\omega_l^{2^L}(u)$ such that $E(e^{q\Omega_s}) =
e^{-(q-\zeta(q)) \ln(s)}$.
More precise results used in the proofs
are stated in Appendix~D in \cite{suppA}.

In this section, we will study the scaling properties of scattering
moments associated
with $X= dM$. Thanks to the discussion in Section~\ref{sconnormals},
one can easily show that all our results
can be extended to $X(t) = M(t) = \int_0^t dM$.
The following theorem characterizes the behavior of normalized first- and
second-order scattering moments of $dM$:

\begin{theorem}
\label{prop2mrm}
Let $dM$ be a Multifractal Random Measure, then
%
\begin{equation}
\label{c1} \forall j <L, \qquad {\widetilde S}\,dM(j) = 1,
\end{equation}
and if
$\zeta(2)>1$ then as long as $j_1,j_2 < L$,
$\widetilde S \,dM(j_1,j_2)$ depends only on $j_1-j_2$ and there exists
$\widetilde K > 0$ such that for each $j_2 \leq L$
%
\begin{equation}
\label{final1}
\lim_{j_1 \rightarrow-\infty} {\widetilde S}\,dM(j_1,j_2)
= \widetilde{K}.
\end{equation}
\end{theorem}

\begin{figure}

\includegraphics{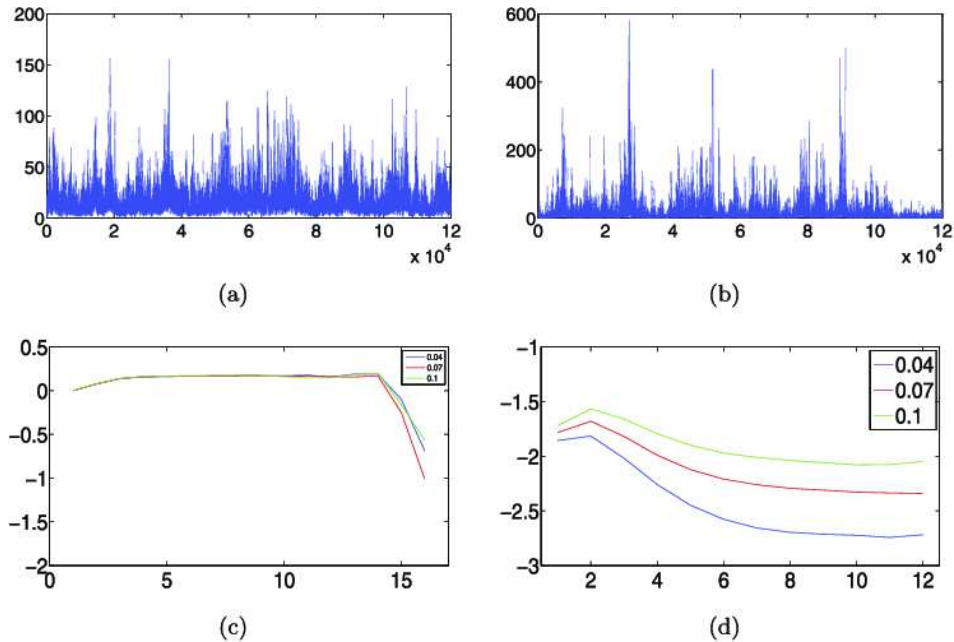}

\caption{\textup{(a)}, \textup{(b)} Realizations\vspace*{1pt} $dM$ of log-normal Multifractal
Random Measures with
$\lambda^2=0.04$ and $\lambda^2=0.1$.
\textup{(c)} $\log_2 {\widetilde S}\,dM(j_1)$
with $\lambda^2=0.04$, $\lambda^2=0.07$ and $\lambda^2=0.1$.
\textup{(d)}\vspace*{1pt} $\log_2 {\widetilde S}\,dM(j_1,j_1+l)$,
for $\lambda^2=0.04$, $\lambda^2=0.07$ and $\lambda^2=0.1$,
as a function of $l$, for $j_1 + l < L$ where $2^L=2^{13}$ is the
integral scale.}
\label{MRMscattfig}
\end{figure}

The proof is in Appendix~E in \cite{suppA}.
Let us illustrate this
theorem in the log-normal case.
Figures~\ref{MRMscattfig}(a), (b) displays two realizations\vspace*{1pt} of
log-Normal MRM cascades for $\lambda^2 = 0.04$
and $\lambda^2 = 0.07$, with an
integral scale $2^L = 2^{13}$. Figure~\ref{MRMscattfig}(c) shows
estimations of
normalized first-order scattering moments for $\lambda^2 = 0.04$,
$\lambda^2 = 0.07$ and $\lambda^2 = 0.1$.
As predicted by Theorem~\ref{prop2mrm},
$\log_2 {\widetilde S}\,dM (j_1) = 0$ for $j_1 < L = 13$.
The second-order scattering moments for the same values of $\lambda^2$ are
displayed in Figure~\ref{MRMscattfig}(d).
As expected from Theorem~\ref{prop2mrm},
$\log_2 {\widetilde S}\,dM (j_1,j_2)$ only depends on $j_2 - j_1$ for
$j_2 < L$.
It converges to a constant\vspace*{1pt} $\widetilde K$ when $j_2-j_1$ increases.

With a Taylor expansion, one can show that, for large $j_2-j_1$,
$\widetilde K$ is a linear function of
$\lambda$ up to some $O(\lambda^2)$ additive term.
This is numerically verified by Monte Carlo simulations which
shows that $\widetilde K \approx0.82 \lambda$. We see here again the
correspondence between scattering coefficients and intermittency
measurements. The constant $0.82$ depends upon the choice of wavelet
$\psi$.

Another important class of stochastic self-similar processes
are the Multi-fractal Random Walks (MRW) \cite{MuDeBa00,BM03}, defined as
$X(t)=B(M(t))$, where $B(t)$ is a standard Brownian motion
and $M(t)$ is an MRM. MRWs are stochastic volatility models,
which account for asset price fluctuations in
financial markets by mimicking the stochastic behavior of asset
volatility \cite{MuDeBa00,BKM06,BKM13}. Appendix~F in
\cite{suppA} shows
that the MRW satisfies the analog of Theorem~\ref{prop2mrm},
with the same second-order behavior, revealing the underlying intermittency
structure introduced by $M(t)$, but different first-order asymptotics,
due to
the scaling of the Brownian motion.

\section{Parametric model estimation with scattering moments}
\label{modelselection}

Section~\ref{scatestimasec} introduces estimators of scattering
moments.
Section~\ref{generamoment} applies
the generalized method of simulated moments
to estimate the parameters of data generating models.
Sections~\ref{turbusection} and \ref{financesection}
analyze the scattering moments of
turbulence data and financial time series to evaluate
fractional Brownian, L\'evy stable and multifracal cascade models.

\subsection{Estimation of scattering moments}
\label{scatestimasec}

We study scattering moment estimators introduced in
\cite{stephane}, and compute upper bounds
of their mean-square error.
A scattering moment
${\overline S}X(j_1,\ldots,j_m) = \mathbf{E}(||X \star\psi_{j_1}| \star
\cdots | \star\psi_{j_m}|)$
is estimated
by replacing the expected
value by a time averaging at a scale $2^M$. It is calculated
with a time window
$\phi_M (t) = 2^{-M} \phi(2^{-M} t)$ with $\int\phi(t)\,dt = 1$.
For any $(j_1,\ldots,j_m) \in{\mathbb{Z}}^m$ with $j_k \leq M$, the
estimator is
%
\begin{equation}
\label{scattrans}
\widehat SX(j_1,\ldots,j_m) = ||X \star
\psi_{j_1}| \star\cdots | \star\psi_{j_m}|\star\phi_M
(t_0),
\end{equation}
where $t_0$ is typically in the middle of the domain where $X(t)$ is known.
Since $\int\phi_M (t)\,dt = 1$, this estimator is unbiased
$\mathbf{E}(\widehat SX(j_1,\ldots,j_m)) = {\overline S}X(j_1,\ldots,j_m)$.
The following theorem, proved in Appendix~G in \cite{suppA}, gives an upper bound of the
mean squared estimation error at each scale.

\begin{theorem}
\label{ggpp}
Suppose that the Fourier transform $\Phi(\omega)$ of $\phi$ satisfies
%
\begin{equation}
\label{phsdfo}
\bigl|\Phi(\omega)\bigr|^2 \leq\frac{1} 2 \sum
_{j=1}^{\infty} \bigl( \bigl|\Psi \bigl(2^j \omega
\bigr)\bigr|^2 + \bigl|\Psi\bigl(- 2^j \omega\bigr)\bigr|^2
\bigr)\qquad\mbox{with }\Phi(0) = 1.
\end{equation}
If $X$ has stationary increments and
$\mathbf{E}(|X \star\psi_{j_1}|^2) < \infty$ then
the mean squared estimation error
\begin{eqnarray*}
\varepsilon(j_1) &\stackrel{\mathrm{def}} {=}& \mathbf{E}\bigl(\bigl|\widehat
SX(j_1) - {\overline S}X(j_1)\bigr|^2\bigr) \\
&&{}+ \sum
_{m=2}^{\infty} \sum
_{-\infty< j_2,\ldots,j_m \leq M} \mathbf{E}\bigl(\bigl|\widehat SX(j_1,\ldots,j_m)
- {\overline S}X(j_1,\ldots,j_m)\bigr|^2\bigr)
\end{eqnarray*}
satisfies
%
\begin{equation}
\label{scattrans2f3f} \varepsilon(j_1) \leq\sigma^2\bigl(|X \star
\psi_{j_1}|\bigr) - \sum_{m=2}^{\infty} \sum
_{-\infty< j_2,\ldots,j_m \leq M} \bigl|{\overline S}X(j_1,\ldots,j_m)\bigr|^2.
\end{equation}
\end{theorem}

When $j_1$ is close to $M$ then
$|X \star\psi_{j_1} (t)|$ decorrelates slowly relatively to the
averaging window scale $2^M$ so $\varepsilon(j_1)$ is large,
but it is
bounded by $\sigma^2(|X \star\psi_{j_1}|)$.
Large variance estimators $\widehat SX(j_1,\ldots,j_m)$
are eliminated by keeping only
small scales $j_k \leq J$ for $1\leq k \leq m$, with
$M-J$ sufficiently large.
For most classes of random processes, including fractional Brownian motions
and multi-fractal random walks,
we observe numerically that
$\varepsilon(j_1)$ converges to zero as the averaging scale
$2^M$ goes to $\infty$. Equation (\ref{scattrans2f3f}) proves that it is
the case if for all $j_1$
\[
\sigma^2\bigl(|X \star\psi_{j_1}|\bigr) = \sum
_{m=2}^{\infty} \sum_{-\infty< j_2,\ldots,j_m \leq\infty}
\bigl|{\overline S}X(j_1,\ldots,j_m)\bigr|^2.
\]
This energy conservation has been conjectured
for large classes of processes in \cite{stephane}, but it is not proved.

For ${n}$ independent realizations $\{ X_k (t)\}_{1 \leq k \leq{n}}$,
we compute an averaged scattering estimator
%
\begin{equation}
\label{ksdfnsdf} \widehat SX = {n}^{-1} \sum
_{k=1}^{n}\widehat SX_k.
\end{equation}
Its variance is thus
reduced by ${n}^{-1}$. When ${n}$ goes to $\infty$,
the central limit theorem proves that
$\widehat SX - {\overline S}X$ converges to a zero-mean normal
distribution whose
variance goes to $0$.

\subsection{Generalized method of simulated scattering moments}
\label{generamoment}

The generalized method of simulated moments \cite{fadden} computes
parameter estimators for
data generative models, from arbitrary families of moments. We apply it to
scattering moments.

Suppose that $\{ X_k \}_{1 \leq k \leq{n}}$ are ${n}$ independent realizations
of a parametric model $Y_\theta$. Then $\widehat SX$ is an unbiased estimator
of ${\overline S}Y_\theta$, so $m(\theta) = \mathbf{E}(\widehat SX) -
{\overline S}Y_\theta= 0$.
The generalized method of moments estimates
this moment condition with
%
\begin{equation}
\label{GMM0} \hat{m}(\theta) = \widehat SX - {\overline
S}Y_\theta.
\end{equation}
When ${n}$ goes to $\infty$ the central limit theorem proves
that $\hat{m}(\theta)$ converges to a normal distribution.
The generalized method of moments finds the parameter $\hat{\theta}$
such that $\hat{\theta}=\mathop{\argmin}_{\theta} \hat{m}(\theta)
W \hat{m}(\theta)^T$
for appropriate matrices $W$. Setting $W = Id$ gives
%
\begin{equation}
\label{GMM00} \hat{\theta}_1 = \mathop{\argmin}_{\theta} \|\widehat SX - {
\overline S}Y_\theta\|^2.
\end{equation}
The two-step generalized method of moment updates the first estimator
$\hat{\theta}_1$ by setting
$W = \widehat{W}_{\hat\theta_1}$, where $\widehat{W}_\theta$
is the inverse of the empirical covariance
%
\begin{equation}
\label{GMM} \widehat{W}_{\theta} = \Biggl( {n}^{-1} \sum
_{k=1}^{{n}} (\widehat{S}X_k - {
\overline S}Y_\theta) (\widehat{S}X_k - {\overline
S}Y_\theta)^T \Biggr)^{-1}.
\end{equation}
It computes
%
\begin{equation}
\label{nomsdfnsdf} \hat{\theta}= \mathop{\argmin}_{\theta} \hat{m}(\theta)
\widehat{W}_{\hat\theta_1} \hat{m}(\theta)^T.
\end{equation}

Since in general we cannot compute ${\overline S}Y_\theta$ analytically,
according to the simulated method of moments \cite{fadden},
${\overline S}Y_\theta$ is replaced in (\ref{GMM0}) and (\ref{GMM})
by an estimator $\widehat SY_{\theta}$
calculated with a Monte Carlo simulation.
This estimator is computed with ${n}' \gg{n}$ realizations which are
adjusted in order to yield a negligible mean-square error
$\mathbf{E}(\|\widehat SY_{\theta} - {\overline S}Y_\theta\|^2)$.
We also compute a $p$-value for the
null hypothesis which supposes that the parameterized model is valid.
The J-test \cite{hansen} is a chi-squared goodness of fit test
normalized by the ${p}-d$ degrees of freedom:
%
\begin{equation}
\label{jvalue} {\chi^2_{\mathrm{red}}} = ({p}- d)^{-1}
{n} \hat{m}(\hat{\theta}) \widehat W_{\hat\theta} \hat m(\hat\theta)^T.
\end{equation}
Under the null hypothesis, $({p}-d)\chi^2_{\mathrm{red}}$ asymptotically follows
a chi-squared distribution with ${p}-d$ degrees of freedom.

In practice, one must optimize
the number ${p}$ of scattering moments
to have enough discriminability with
an estimator having small variance.
In the present work, we limit ourselves to first- and second-order scattering:
\[
{\overline S}X = \bigl( {\overline S}X (j_1), {\overline
S}X(j_1,j_2) \bigr)_{J_0 < j_1 \leq J, j_1 < j_2 \leq J}.
\]
Indeed, finest scale coefficients $j_1 \leq J_0$ are removed to avoid errors
due to aliasing, discretization or to some data smoothing,
Similarly, large scale coefficients \mbox{$j_i > J$} are also removed to
since to the largest variance estimators, and
${\overline S}X(j_1,j_2)$ for $j_2 \leq j_1$ are also eliminated because
they carry little information on $X$. The resulting scattering vector
has $J-J_0$ first-order scattering moments and $(J-J_0-1)(J-J_0)/2$
second-order moments.

%
%

If we observe a single realization $X(t)$ where sufficiently far away
wavelet coefficients become independent, 
$\widehat SX$ becomes asymptotically normal when computed
at intervals $\Delta$:
\begin{eqnarray}
\widehat SX_k (j_1) &=& |X \star
\psi_{j_1}|\star\phi_M (k \Delta )\quad\mbox{and}
\nonumber
\\[-8pt]
\label{getindeprealizations}\\[-8pt]
\nonumber
\widehat
SX_k (j_1,j_2) &=& \bigl||X \star
\psi_{j_1}|\star\psi_{j_2}\bigr| \star\phi_M (k \Delta).
\end{eqnarray}
The goodness of fit J-test
supposes that the variable (\ref{jvalue}) follows a chi-squared distribution
with $p-d$ degrees of freedom,
which requires that the estimators $\widehat SX_k$ are independent
for different $k$.
If $X$ has an integral scale $T$, as in multifractal cascades, then
increments are independent at distances larger than $T$. One can thus
set $\Delta= 2 T$. Other processes, such as fractional
Brownian motions have no integral scales but their wavelet coefficients become
nearly independent at distances much larger than the scale. Nearly
independent estimators are thus obtained if $\Delta\gg2^M$.

The situation is easier if we are only interested in
the parameter estimator $\hat\theta$ with (\ref{nomsdfnsdf}),
without goodness of fit.
Its consistency requires that $\widehat SX$ converges to a normal
distribution, but we can
estimate its covariance up to an unknown multiplicative factor.
Thus, setting $\Delta= 1$ only
introduces a multiplicative factor in the covariance estimation,
which does not affect the estimator $\hat\theta$\vspace*{-9pt} in (\ref{nomsdfnsdf}).

%

\subsection{Intermittency estimation on multiplicative cascades}
The properties of the Scattering Method of Moments are
illustrated on the estimation of the intermittency parameter
$\theta= \lambda^2$ for multifractal random measures.
Section~\ref{scattMRMsection} proves that normalized second-order
scattering moments converge to a constant $\widetilde K$ which is
proportional to $\lambda$, showing that
the intermittency $\lambda^2$ is characterized
by first- and second-order scattering moments.
However, the information is not just
carried by this asymptotic value, which is why all scattering moments
are used for the estimation. The scattering estimation is
compared with two estimators dedicated to
this particular estimation problem \cite{BKM13}.

Scattering moment estimators are computed
from ${n}$ independent realizations of size $2^{11}$
of a multifractal random measure
having an integral scale $T = 2^{10}$.
The total number of data points is $N= {n}\cdot2^{11}$,
%
%
and we set $J_0 = 0$.
For different values of $N = {n}\cdot2^{11}$,
we report in Table~\ref{intermittencytable}
the value of $J$ which minimizes the mean squared error
$\mathbf{E}(|\hat{\theta} - \theta|^2)$, estimated with Monte Carlo
simulations.
We also give the average value of
the reduced $\chi_{\mathrm{red}}^2$ test\vspace*{2pt} in (\ref{jvalue}) and the
model $p$-value.
For small values of ${n}$, the covariance
of $\widehat SX$ is computed up to a multiplicative constant, from
correlated scattering coefficients calculated within each realization
with $\Delta= 1$ in~(\ref{getindeprealizations}).
It leads to a good estimation of $\hat\theta$ but the
model $p$-value cannot be estimated.

\begin{table}
\tabcolsep=0pt
\tablewidth=\textwidth
\caption{Estimation of $\lambda^2$
for a multi-fractal random measure.
The table gives
the mean and the standard deviation of estimators computed with
a wavelet moment regressions (\protect\ref{moonsdfs}),
a log covariance regression (\protect\ref{moonsdfs2}),
and the method of simulated scattering moments,
for several values of $\lambda^2$ and
several sample sizes $N$}
\label{intermittencytable}
\begin{tabular*}{\tablewidth}{@{\extracolsep{\fill}}lcd{5.9}d{5.10}d{5.9}ccc@{}}
\hline
$\bolds{\lambda}^{\mathbf{2}}$
& \multicolumn{1}{c}{$\bolds{N}$}
& \multicolumn{1}{c}{$\hat{\bolds{\theta}}$ \textbf{wavelet}}
& \multicolumn{1}{c}{$\hat{\bolds{\theta}}$ \textbf{log-cov}}
& \multicolumn{1}{c}{$\hat{\bolds{\theta}}$ \textbf{scattering}}
& \multicolumn{1}{c}{$\bolds{J}$}
& \multicolumn{1}{c}{$\bolds{\chi}^{\mathbf{2}}_{\mathbf{red}}$}
& \multicolumn{1}{c@{}}{$\bolds{p}$\textbf{-value}}\\
\hline
$0.02$
& $10^6$
& 0.025 (\pm 2 \cdot 10^{-3}
& 0.02 (\pm 2 \cdot 10^{-4}
& 0.02 (\pm 2 \cdot 10^{-4}
& $7$
& $1.1 \pm 0.3$
& $0.7 \pm 0.3$ \\
$0.05$
& $10^6$
& 0.055 (\pm 2 \cdot10^{-3}
& 0.05 (\pm 6 \cdot10^{-4}
& 0.05 (\pm 3 \cdot10^{-4}
& $6$
& $0.8 \pm 0.3$
& $0.5 \pm 0.3$ \\
$0.1$
& $10^6$
& 0.105 (\pm 4 \cdot10^{-3}
& 0.1 (\pm 10^{-3}
& 0.1 (\pm 10^{-3}
& $5$
& $0.8 \pm 0.5$
& $0.5 \pm 0.3$ \\
$0.1$
& $10^5$
& 0.109 (\pm 10^{-2}
&0.1 (\pm3 \cdot10^{-3}
&0.1 (\pm2 \cdot10^{-3}
&$5$
&$0.7 \pm0.3$
&$0.3 \pm0.3$\\
$0.1$
& $10^4$
& 0.12 (\pm3 \cdot10^{-2}
& 0.1 (\pm1.3 \cdot10^{-2}
& 0.1 (\pm9 \cdot10^{-3}
&$5$
&N$/$A
& N$/$A\\
\hline
\end{tabular*}
\end{table}

The intermittency parameter of multifractal random measures can also be
estimated directly from wavelet coefficients.
Section~\ref{scattMRMsection} explains that
the scaling exponent of wavelet moments of order $q$ is
$\zeta(q) =  (\frac{1}{2}+\lambda^2 )q - \frac{\lambda
^2}{2} q^2$.
It results that $ \lambda^2 = 2 \zeta(1) - \zeta(2)$. The
intermittency parameter can thus
be estimated with a linear regression on the estimated
first- and second-order moments of wavelet coefficients at\vspace*{-3pt} scales $2^j
< 2^L$:
%
\begin{equation}
\label{moonsdfs}
2 \log_2 \mathbf{E}\bigl(|X \star
\psi_j|^2\bigr) - \log_2 {\mathbf{E}\bigl(| X \star
\psi_j|\bigr)^2 \approx{j\bigl(\zeta(2) - 2 \zeta(1)\bigr)}} +
C.
\end{equation}
The wavelet moments $\mathbf{E}(|X \star\psi_j|^2)$ and $\mathbf{E}(| X
\star\psi_j|)$
are estimated with empirical averages of
$|X \star\psi_j|$ and $| X \star\psi_j|^2$, calculated
from the $N$ data samples.
An improved estimator has been introduced
in \cite{BKM06,BKM13} with a regression
on the covariance of the log of the
multifractal random measure. One can indeed prove\vspace*{-3pt} that
%
\begin{equation}
\label{moonsdfs2} \operatorname{Cov} \bigl(\log\bigl| X \star\psi_j(t)\bigr|,\log\bigl| X
\star\psi _j(t+l)\bigr| \bigr) \simeq-\lambda^2 \ln \biggl(
\frac{l}{2^L} \biggr)+ o \biggl(\frac{j} l \biggr),\vspace*{-3pt}
\end{equation}
which leads to lower variance estimations.

Table~\ref{intermittencytable} shows that
the scattering moment estimation of $\lambda^2$
has a smaller variance than the regression of first- and second-order wavelet
moments. This is due to the low variance of the scattering estimators which
are computed with nonexpansive operators.
It gives comparable results
with the log-covariance estimator, which was optimized for this
problem \cite{BKM13}, and which is also closely related
to the wavelet leaders estimator of \cite{abryleaders}.
In the case of log-normal multiplicative cascades, the log-covariance
estimator does not suffer from outliers, although\vspace*{1pt} it is not
a contractive operator, and the log-log linear regression
used to recover $\hat{\lambda}^2$ in~(\ref{moonsdfs2}) reduces
the variance of the estimator. This explains the similar
behavior with respect to the GMM scattering estimator.

The J-test validates
the multifractal model, since we obtain a normalized J-test
with mean and standard deviation close to $1 \pm\sqrt{\frac{2}{{p}-1}}$,
corresponding to mean and standard deviation of a chi-squared
distribution with ${p}-1$ degrees of freedom.
The resulting $p$-values for rejecting the true model are of the order of $0.5$.
As expected, reducing the maximum scattering scale $J$ improves
the estimation of $\lambda^2$ for high intermittencies. It removes large
variance coefficients. However, numerical experiments confirm
that the generalized method of moments is robust to the choice of $J$,
because the inverse covariance $\widehat W$ in (\ref{nomsdfnsdf})
reduces the
impact of high variance coefficients.

\subsection{Estimation of Blumenthal--Gethoor index on L\'{e}vy processes}

We apply the same methodology to estimate the Blumenthal--Gethoor index
of L\'{e}vy processes, defined as
\[
\beta= \inf \biggl\{ r \geq0 \mbox{ s.t. } \int_{|x| \leq1}
|x|^r\, d\Pi(x) < \infty \biggr\},
\]
where $\Pi(x)$ is the L\'{e}vy measure associated to an observed L\'
{e}vy process $X(t)$.
If $X(t)$ is $\alpha$-stable, then $\beta= \alpha$.
This index can be estimated using spectral methods in \cite
{belomestny2010spectral}, which require
an estimation of the characteristic function.

We concentrate in the case of $\alpha$-stable processes, and
we assume here that $1 < \alpha\leq2$, which implies that we cannot consider
the covariance of $\widehat SX$. Instead, we use the simplified GMM estimator
(\ref{GMM00}). Besides the simplified
GMM scattering estimator, we also consider
a log-linear regression on wavelet coefficients
\[
\log_2 \mathbf{E}|X \star\psi_j| \simeq
\alpha^{-1} j,
\]
and also a log-linear regression on scattering moments, using
the asymptotic results of Section~\ref{Levy}:
\[
\log_2 SX(j) \simeq\alpha^{-1} j,\qquad\log_2
\widetilde{S}X(j_1,j_1+l) \simeq\bigl(\alpha^{-1}-1
\bigr) l.
\]
For that purpose, we estimate scattering moments
$\widehat{S}X(j_1,j_2)$ with $j_2\geq j_1 + \delta$ and $\delta=3$.
Table~\ref{intermittencytablelevy} shows that
a regression on scattering moments improves the performance
of wavelet regression, and that GMM scattering significantly
improves with respect to the regression method.
Second-order scattering coefficients produce
statistically significant new information, which improves
regression results. Moreover, the information on $\alpha$
is not just contained in the asymptotic regime $j_2 \gg j_1$ but also
in the
transient regime, which is exploited by the GMM estimator.

\begin{table}
\tablewidth=\textwidth
\caption{Estimation of $\alpha$
for $\alpha$-stable L\'{e}vy.
The table gives
the mean and the standard deviation of estimators computed with
a wavelet moment regressions (\protect\ref{moonsdfs})
and the method of simulated scattering moments,
for several values of $\alpha$ for $N= 2^{20}$}
\label{intermittencytablelevy}
\begin{tabular*}{\tablewidth}{@{\extracolsep{\fill}}lcd{4.8}d{3.8}c@{}}
\hline
$\bolds{\alpha}$
& \multicolumn{1}{c}{$\hat{\bolds{\alpha}}$ \textbf{wavelet} \textbf{regr.}}
& \multicolumn{1}{c}{$\hat{\bolds{\alpha}}$ \textbf{scatt.} \textbf{regr.}}
& \multicolumn{1}{c}{$\hat{\bolds{\alpha}}$ \textbf{scattering} \textbf{GMM}}
& $\bolds{J}$ \\
\hline
$1.1$
&  $1.2\pm0.1$\phantom{000}
& 1.2(\pm8 \cdot 10^{-2}
& 1.1(\pm4 \cdot 10^{-2}
& $7$ \\
$1.5$
& $1.55\pm 7 \cdot10^{-2}$
& 1.51(\pm 6 \cdot10^{-2}
& 1.5(\pm 2 \cdot10^{-2}
& $5$ \\
\hline
\end{tabular*}
\end{table}

\subsection{Turbulence energy dissipation}
\label{turbusection}
Turbulent regimes appear in a wide
variety of experimental situations, and are characterized by random fluctuations
over a wide range of time and space scales.
Making a theory of the famous Richardson ``energy cascade'' across the
inertial range
remains one of the main challenges in classical physics \cite{BF}.
Normalized scattering moments are computed over
dissipative measurements of a turbulent gas,
to analyze their self-similarity and intermittency
properties.
This study does not pretend evaluating general turbulence physical models.
However, it shows that one can have confident model evaluations
from data sets, despite intermittency phenomena.
\setcounter{footnote}{1}

\begin{figure}

\includegraphics{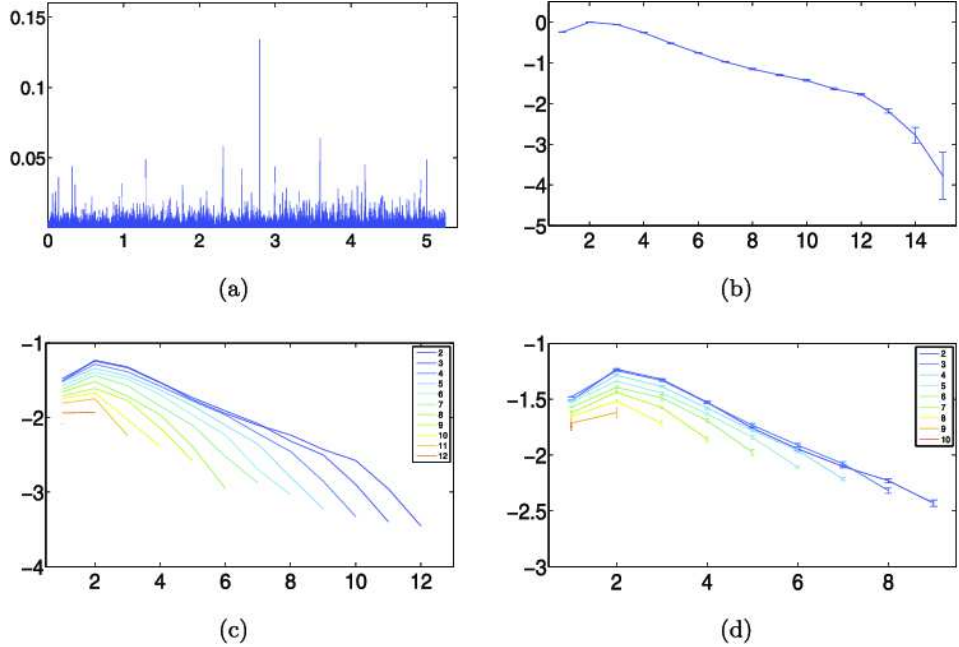}

\caption{\textup{(a)} Realization of dissipation $X(t) =  (\frac{\partial
v}{\partial t}  )^2$ in a turbulent flow.
\textup{(b)} Estimation $\log_2 \widehat{\widetilde S}X (j_1) $
as a function of $j_1$, calculated from $4$ realizations
of $2^{19}$ samples each.
\textup{(c)} $\log_2 \widehat{\widetilde S}X(j_1,j_1+l)$
as a function of $l$, for $2 \leq j_1 \leq12$.
\textup{(d)} $\log_2 \widehat{\widetilde S}X(j_1,j_1+l)$ in the inertial range
$j_1 + l < L -1=10$. We plot the
confidence intervals
corresponding to the standard deviation of the estimated $\log_2
\widehat{\widetilde S}X(j_1,j_1+l)$.}
\label{turbulencefigure1}
\end{figure}

The data we used has been recorded by the group
of B. Castaing in Grenoble in a low temperature
gazeous Helium jet in which the Taylor scale-based Reynolds
number is $R_\lambda= 703$ \cite{chan00}.
A single probe provides measures of
velocity temporal variations at a fixed space location that involve
both Lagrangian and Eulerian
fluctuations.
%
Figure~\ref{turbulencefigure1}(a) shows
a sample of the surrogate dissipation field
$X(t)$ as a function of time,
estimated from the experimental velocity records.\footnote{One assumes
the validity of the Taylor
hypothesis \cite{BF}.}
The Kolmogorov (dissipative) scale $\eta$ is observed at approximately
$2^2$ sample points, whereas the
integral scale is approximately $2^L = 2^{11}$ sample points.

First-order scattering coefficients are normalized at the finest
scale defined by $j_1 = 2$.
These coefficients are
displayed in Figure~\ref{turbulencefigure1}(b).
In the inertial\vspace*{1pt} range
$2^1 = 2^{J_0} < 2^{j_1} \leq2^{L} = 2^{11}$
the scaling law of the exponents is ${\widetilde S}X(j_1) \simeq
2^{-0.25 j_1}$.
If $2^{j_1} \geq2^{L}$ then
${\widetilde S}X(j_1) \simeq2^{-j_1 /2}$ because the low frequencies
of a
turbulent flow becomes Gaussian and independent beyond
the integral scale.

Figure~\ref{turbulencefigure1}(c)
gives estimated normalized second-order coefficients\break $\log_2
\widetilde S X(j_1, j_1+l)$ as a function of $l$,
for different $j_1$.
For $j_2 = j_1+l > L$, the slopes increase up to $-1/2$ because beyond the
integral scale, wavelet coefficients
converge to Gaussian random processes.
Below the integral scale, $j_2 = j_1+l < L-1$,
Figure~\ref{turbulencefigure1}(d)
shows the curves $\log_2 \widetilde S X(j_1,j_1+l)$ with
error bars giving the standard
deviations of each estimated values.
In this inertial range, the average slope of all curves is $-0.2$.
This slope is very different from the $-1/2$ decay of Gaussian processes,
which indicate the presence of intermittent phenomena.
Although these curves are similar,
one can observe that they differ significantly compared to the error bars,
which indicates the self-similarity of turbulence data is violated.
This nonself-similarity is likely to originate from
the fact that, as already observed in \cite{DelMuArn01,castaing02},
Taylor hypothesis does not rigorously hold.
%
%
We consider the three following models for inertial range turbulence:
(i) the square of a Fractional Gaussian Noise parameterized
by $\theta=H$, (ii) the square of the increments of $\alpha$-stable
L\'{e}vy
processes, parameterized by $\theta=\alpha$ and finally  (iii)
log-normal multifractal random measures, parameterized by
the intermittency parameter $\theta=\lambda^2$.
Setting $J_0 = 1$ eliminates
coefficients below the diffusion scale.
We have $N = 4  \cdot10^{6}$ data samples, divided into $4$ realizations.
Within each realization, since
the integral scale is $T = 2 \cdot10^3$, samples
are independent at a distance larger than $T$.
The maximum scale is set to $2^J = 2^8$ but its modification has a marginal
impact on the estimation. The size of the resulting scattering vector is
${p}= (J-J_0+1)(J-J_0)/2 = 28$.

\begin{table}
\tablewidth=220pt
\caption{Parameter estimation
for the turbulence data in Figure \protect\ref{turbulencefigure1}\textup{(a)}, calculated
from Fractional Brownian Noise measures (FBN),
L\'evy stable measures (LS), and Multifractal Random Measures (MRM)}
\label{turublmod}
\begin{tabular*}{220pt}{@{\extracolsep{\fill}}lccc@{}}
\hline
& \textbf{FBN} & \textbf{LS} &\textbf{MRM} \\
\hline
$\hat\theta$ & $H = 0.9$ & $\alpha= 1.98$ & $\lambda^2 = 0.09$ \\
$\chi^2_{\mathrm{red}}$ & $28$ & $29 $ & $29$ \\
$p$-value & ${<}10^{-6}$ &${<}10^{-6}$ & ${<}10^{-6}$ \\
\hline
\end{tabular*}
\end{table}

Table~\ref{turublmod} gives an optimal parameter $\hat\theta$
as well as the value of the $\chi_{\mathrm{red}}^2$
goodness of fit test in (\ref{jvalue}), together with
its $p$-value. All models are rejected with very high confidence.
For nearly the same number of data values,
with integral scales of same size,
Table~\ref{intermittencytable} gave much higher $p$-values
for valid multifractal random measure models of same intermittency.
Fractional Brownian motions
can explain the behavior of first-order scattering,
but not the second-order ($-1/2$ as opposed to $-0.2$). L\'{e}vy stable
processes
have first- and second-order scattering which decay
with a slope of $\alpha^{-1}-1$. To match the slopes in
Figure~\ref{turbulencefigure1}(c), (d),
respectively equal to $-0.25$ and $-0.2$, we would need that
$\alpha\approx1.2$ which is far from the value $\alpha= 1.98$
obtained in
Table~\ref{turublmod}.
Multifractal random measure model misfit comes from their
first-order coefficients which remain constant
whereas turbulence data coefficients decay with a slope close to $-0.2$.

\subsection{Financial time-series analysis}
\label{financesection}

In the following, we analyze the normalized scattering moments
of two financial time series: high-frequency
Euro-Bund trade data\footnote{Euro-Bund is one of the most actively
traded financial asset in the world.
It corresponds to a future contract on an interest rate
of the Euro-zone.} and intraday S\&P 100 index trade data.
Each trade occurs at a given price, whose logarithm is noted $X(t)$.
\begin{figure}

\includegraphics{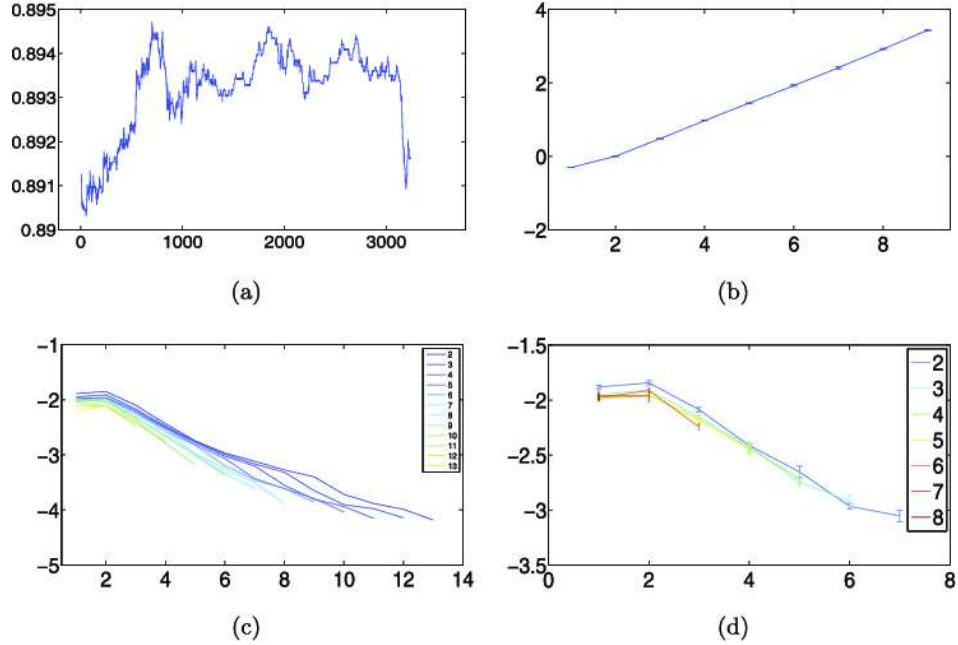}

\caption{\textup{(a)} One day of the deseasonalized Euro-Bund log-price $X(t)$.
\textup{(b)} Estimated $\log_2 \widehat{\widetilde S}X(j_1)$.
\textup{(c)} Estimated $\log_2 \widehat{\widetilde S}X(j_1,j_1+l)$.
\textup{(d)} Estimated $\log_2 \widehat{\widetilde S}X(j_1,j_1+l)$ for $j_1 +
l < 9$.}
\label{bundfig}
\end{figure}

Every single day, the logarithmic returns of the price [i.e., the
increments of $X(t)$]
are computed on rolling $10$ second intervals, after preprocessing the
microstructure noise using the technique advocated in \cite{RR}.
Each day corresponds to $9$ hours of trading, and hence $3240$ increments.
Intraday financial data are subject to strong seasonal
intraday effects.
These effects are removed with
a standard ``deseasonalizing''
algorithm which normalizes the returns
by the square root of the intraday seasonal variance.

Figure~\ref{bundfig} displays the resulting deseasonalized Bund
log-price $X(t)$
and its estimated scattering moments.
The decay of first-order scattering moments is $\log_2 {\widetilde
S}X(j) \sim0.48 j$,
whereas for second-order is
$\log_2 \widehat{\widetilde S}X(j_1,j_1+l) \sim-0.2 l$
for all $j_2 = j_1+l$.
Contrarily to turbulence data, we do not see
an integral scale, beyond which second-order coefficients would have a fast
decay of $-0.5 l$. This is not surprising since the integral scale is
known to be larger than few months \cite{BKM13}.
Figure~\ref{bundfig}(d) gives intra-day second-order coefficients
$j_2 = j_1 + l < 9$. The variance of
$\widehat{\widetilde S}X(j_1,j_1+l)$ is indicated with vertical error bars.
Observe that $\log_2 \widehat{\widetilde S}X(j_1,j_1+l)$
has small variations as a function of $j_1$,
which is a strong indication of self-similarity.

\begin{figure}

\includegraphics{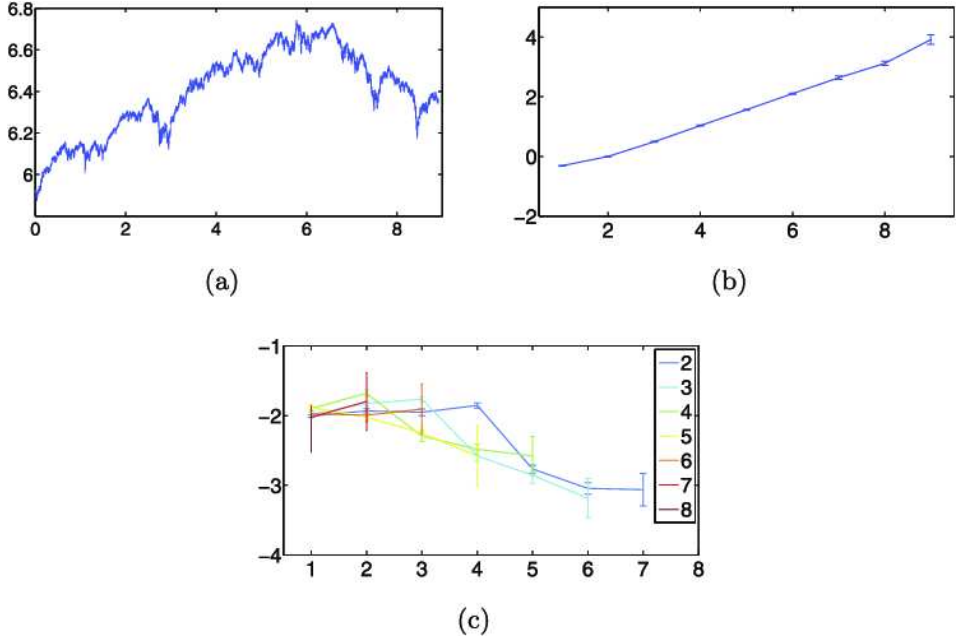}

\caption{\textup{(a)} Three years of the deseasonalized S\&P 100 index
log-price $X(t)$. \textup{(b)} Estimated $ \log_2 \widehat{\widetilde S}X(j_1)$.
\textup{(c)} Estimated $\log_2 \widehat{\widetilde S}X(j_1,j_1+l)$ for $j_1 +
l < 9$.}
\label{spfig}
\end{figure}

The same scattering computations
are performed on the S\&P 100 index,
sampled every $5$ minutes from April 8th, 1997, to December 17th, 2001,
to yield $78$ samples every day.
Figure~\ref{spfig} shows the scattering moments estimated on the S\&P
time series.
Observe that $\log_2 {\widetilde S}X(j_1)$ remains regular for $j_1$
close to $6$,
and that for $j_2 = j_1 + l = 6$ the coefficient $\log_2 {\widetilde
S}X(j_1,j_1+l)$ is
higher than expected, relatively to other coefficients, which means a higher
level of intermittency. $2^6$ roughly corresponds to a trading day $78
\simeq2^6$.

We consider the three following models: (i) fractional Brownian
motions models with
$\theta= H$, (ii) L\'{e}vy stable processes parameterized
with $\theta= \alpha$ and (iii) multifractal random walks
with $\theta= \lambda^2$.
For each model family, Table~\ref{financemod} estimates an optimal
parameter $\hat\theta$ from first- and second-order scattering.
They are computed from a total of $N = 3  \cdot10^6$ (resp., $N =
10^5$) samples for the Euro-Bund (resp., S\&P 100).
The maximum scale $2^J$ is
adjusted to $J=8$ (resp., $J=6$), and we set $J_0 = 1$ to eliminate
discretization effects in both cases.
For fractional Brownian motions, the estimated parameter
$\hat\theta= H = 0.5$
corresponds to a Brownian motion. Brownian motion models explain the
power-spectrum decay of these processes but are known not to be
appropriate because they do not take into account the intermittency
behavior of
financial markets. This
appears in the second-order scattering moments
of Figure~\ref{bundfig}(d) and \ref{spfig}(c), which
have a much slower decay than Brownian motions.
The L\'{e}vy-stable parameters $\alpha$ in
Table~\ref{financemod} are close to $2$ (order 2 moment of financial
time-series are known to be finite).
Estimated models of multifractal random walks show the
existence of intermittency which is larger for the S\&P 100 data set
than for
the Euro-Bund data.
For each model, Table~\ref{financemod} gives the value of
the J-test variable $\chi^2_{\mathrm{red}}$ computed with (\ref{jvalue}).
Multifractal random walks have the lowest value
$\chi^2_{\mathrm{red}}$, and hence provide the better stochastic model of
the data.
However, one cannot
compute a $p$-value because the empirical covariance matrix is
computed from correlated
scattering estimators $\widehat SX_k$ in (\ref{getindeprealizations}).
Because the integral scale is too large, one cannot fix an interval
$\Delta$ providing independent scattering values.

\begin{table}
\caption{The\vspace*{1pt} left and right parts of the table correspond
to Euro-Bund and S\&P 100 time series.
The first row gives the estimated parameter value $\hat\theta$
for Fractional Brownian Motion (FBM),
L\'{e}vy stable processes (LS) and Multifractal Random Walks (MRW)}
\label{financemod}
\begin{tabular*}{\tablewidth}{@{\extracolsep{\fill}}lcccccc@{}}
\hline
& \multicolumn{3}{c}{\textbf{Euro-Bund}} & \multicolumn{3}{c@{}}{\textbf{S\&P}} \\[-4pt]
& \multicolumn{3}{c}{\hrulefill} & \multicolumn{3}{l@{}}{\hrulefill} \\
$\bolds{S}$ & \textbf{FBM} & \textbf{LS} & \textbf{MRW} & \textbf{FBM} & \textbf{LS} & \textbf{MRW} \\
\hline
$\hat\theta$ & $H=0.5$ & $\alpha= 1.95$ & $\lambda^2 = 0.03$ & $H =
0.5$ & $\alpha=1.8 $ & $\lambda^2=0.08$ \\
$\chi^2_{\mathrm{red}}$ & $29$ & $26$ & $23$ & $17$ & $16$ & $10$\\
\hline
\end{tabular*}
\end{table}


\begin{supplement}[id=suppA]
\stitle{Proofs of theorems}
\slink[doi]{10.1214/14-AOS1276SUPP}
\sdatatype{.pdf}
\sfilename{aos1276\_supp.pdf}
\sdescription{We provide the technical derivations of all the results.}
\end{supplement}

\printaddresses
\end{document}